\documentclass[aps,prd,superscriptaddress,twocolumn,nofootinbib]{revtex4-2}

\usepackage[utf8]{inputenc}
\usepackage[T1]{fontenc}

\usepackage{mathtools}
\usepackage{amssymb}
\allowdisplaybreaks[4]

\usepackage{graphicx}

\usepackage{xcolor}
\definecolor{navyblue}{rgb}{0.0,0.0,0.5}
\definecolor{olive}{rgb}{0.5,0.4,0.0}

\usepackage{xparse}
\NewDocumentCommand{\dv}{m g}{%
	\IfNoValueTF{#2}{\frac{\mathrm{d}}{\mathrm{d}#1}}{\frac{\mathrm{d}#1}{\mathrm{d}#2}}%
}
\NewDocumentCommand{\pdv}{m g}{%
	\IfNoValueTF{#2}{\frac{\partial}{\partial #1}}{\frac{\partial #1}{\partial #2}}%
}
\NewDocumentCommand{\order}{m}{\mathcal{O}\!\left(#1\right)}

\newcommand{\dd}{\mathrm{d}}

\usepackage{xcolor}

\usepackage{hyperref}
\hypersetup{
	colorlinks=true,
	citecolor=orange,
	filecolor=navyblue,
	linkcolor=navyblue,
	urlcolor=navyblue
}

\usepackage{lipsum}
\usepackage{tensor}
\usepackage{physics}
\usepackage{comment}

\makeatletter
\renewcommand\onecolumngrid{%
	\do@columngrid{one}{\@ne}%
	\def\set@footnotewidth{\onecolumngrid}%
	\def\footnoterule{\kern-6pt\hrule width 1.5in\kern6pt}%
}
\renewcommand\twocolumngrid{%
	\def\footnoterule{%
		\dimen@\skip\footins\divide\dimen@\thr@@
		\kern-\dimen@\hrule width.5in\kern\dimen@}%
	\do@columngrid{mlt}{\tw@}%
}
\makeatother

\begin{document}
	\title{The \emph{quasi-topological} space of quasi-topological gravities}
	
	\author{Pablo Bueno}
	\email{pablobueno@ub.edu}
	\affiliation{Departament de Física Quàntica i Astrofísica, Institut de Ciències del Cosmos, Universitat de Barcelona, Martí i Franquès, 1, 08028, Barcelona, Spain}
	
	\author{Robie A. Hennigar}
	\email{robie.a.hennigar@durham.ac.uk}
	\affiliation{Centre for Particle Theory, Department of Mathematical Sciences, Durham University, Durham DH1 3LE, United Kingdom}
	
	\author{Matteo G. Marrone}
	\email{matteomarrone@icc.ub.edu}
	\affiliation{Departament de Física Quàntica i Astrofísica, Institut de Ciències del Cosmos, Universitat de Barcelona, Martí i Franquès, 1, 08028, Barcelona, Spain}
	
	\author{Ángel J. Murcia}
	\email{ajm386@cam.ac.uk}
	\affiliation{Department of Applied Mathematics and Theoretical Physics, University of Cambridge, Wilberforce Road, Cambridge CB3 0WA, United Kingdom}

	\begin{abstract}

		We investigate general properties of quasi-topological theories of gravity involving infinite towers of polynomial densities and of their regular vacuum black hole solutions. We show that every positive-mass regular vacuum solution necessarily possesses a de Sitter core, while Minkowski and anti-de Sitter cores are excluded. Regarding the horizon structure, we prove that asymptotically flat and anti-de Sitter regular black holes always admit extremal configurations. We find that there is no upper bound on the number of horizons and that horizons of arbitrarily high degeneracy can arise. These results extend to more general theories of gravity with second-order equations on spherically symmetric backgrounds. Finally, we characterize the space of quasi-topological theories admitting regular black holes, showing that it is contractible (in particular, path-connected and simply connected) and that regularity is stable under suitably small deformations of the higher-order couplings.  In the sense of Spanier, this space is naturally a contractible \emph{quasi-topological} space. 
		Additionally, we prove that any theory defined by
		a set of couplings up to any arbitrarily high,
		but finite, curvature order can be completed into a theory yielding regular black hole solutions, while
		preserving the same set of lower-order couplings.
		These results show that regular solutions arise in a broad class of quasi-topological gravities and do not require fine-tuning of the couplings.
		
	\end{abstract}

	\maketitle

	\section{Introduction}
	
	The singularity theorems of Penrose and Hawking~\cite{Penrose:1964wq,Hawking:1970zqf,Hawking:1973uf} established that, under rather general assumptions, general relativity (GR) predicts the formation of spacetime singularities, where curvature invariants become unbounded and the classical description of spacetime eventually breaks down. Understanding how these singularities are resolved is therefore one of the central open problems in gravitational physics.
	
	One possibility is that the classical geometric description of spacetime itself ceases to be meaningful in the high-curvature regime, and that a complete quantum theory of gravity is required to describe the black hole interior. A different possibility, which motivates the regular black hole (RBH) program \cite{Carballo-Rubio:2018pmi,Carballo-Rubio:2022kad,Carballo-Rubio:2025fnc,Carballo-Rubio:2026gwg}, is that the metric description remains useful beyond the regime of validity of GR, while the singularity is removed by suitable modifications of the gravitational dynamics. From this perspective, one searches for black hole geometries which reproduce the usual behavior at sufficiently large distances but replace the singular interior by a regular region with bounded curvature. 
	
	This idea has a long history. Some of the earliest and best-known phenomenological examples are the so-called Bardeen, Hayward and Dymnikova geometries~\cite{Sakharov:1966aja,Bardeen1968,Poisson:1988wc,Hayward:2005gi,Dymnikova:1992}. RBHs replace the singular center by a regular core, with bounded curvature. These geometries are usually studied within GR by introducing an effective stress-energy tensor capable of supporting the corresponding regular spacetime. From this perspective, the regularization of the black hole interior is attributed to a suitable matter source rather than to a modification of the gravitational dynamics themselves. However, in order to address dynamical questions---such as their formation, stability, or the fate of their inner horizons---it is important to identify gravitational theories for which regular black holes appear as solutions rather than being imposed at the level of the metric.
	
	A widely explored realization is provided by GR coupled to nonlinear electrodynamics (NLED). Suitable choices of the electromagnetic Lagrangian can support regular black hole geometries, although typically only on special loci of the space of solutions
	\cite{Ayon-Beato:1998hmi,Bronnikov:2000vy,Ayon-Beato:2000mjt,Bronnikov:2000yz,Ayon-Beato:2004ywd,Dymnikova:2004zc,Berej:2006cc,Balart:2014jia,Fan:2016rih,Bronnikov:2017sgg,Junior:2023ixh,Alencar:2024yvh,Bronnikov:2024izh,Bolokhov:2024sdy,Skvortsova:2024wly,Murk:2024nod,Li:2024rbw,Zhang:2024ljd}.
	These constructions demonstrate explicitly that spacetime singularities can be avoided within a classical field-theoretic framework. At the same time, the regularization mechanism relies on an important degree of fine-tuning, both at the level of the theory and at the level of the solutions. In addition, when the matter sector is removed, the usual singular vacuum solutions of GR remain. Thus, while GR coupled to appropriate matter provides a realization of the RBH program, it does not by itself modify the singular vacuum sector of the gravitational theory.\footnote{It has also been possible to identify higher-derivative extensions of Einstein-Maxwell theory which feature charged regular black hole solutions \cite{Cano:2020ezi,Cano:2020qhy,Bueno:2021krl,Cano:2022ord,Bueno:2025dqk} for arbitrary values of the (nonzero) charge and mass. However, as soon as charge is set to zero, a singular solution is recovered. Alternative mechanisms leading to RBHs have been proposed \emph{e.g.,} in ~\cite{Frolov:1989pf,Barrabes:1995nk,Bonanno:2000ep,Nicolini:2005vd,Greenwood:2008ht,Olmo:2012nx,Saini:2014qpa,Frolov:2014jva,Balakin:2015gpq,Bazeia:2015uia,Chamseddine:2016ktu,Bambi:2016xme,Bejarano:2017fgz,Kawai:2017txu,Cano:2018aod,Guerrero:2020uhn,Brandenberger:2021jqs,Olmo:2022cui,Junior:2024xmm,Harada:2025cwd}. }
	
	A different mechanism has recently been found within higher-curvature gravity. In~\cite{Bueno:2024dgm}, it was shown that the Schwarzschild singularity can generically be resolved in any dimension $D\geq 5$ when an infinite tower of higher-order polynomial curvature corrections is added to the Einstein-Hilbert action. The theories studied fall into the category of quasi-topological (QT) gravities~\cite{Oliva:2010eb,Myers:2010ru,Dehghani:2011vu,Ahmed:2017jod,Cisterna:2017umf,Bueno:2019ycr,Bueno:2022res,Moreno:2023rfl,Bueno:2025qjk,Bueno:2026oyg,Murcia:2026pmg}.
	These are characterized by possessing second-order field equations on spherically symmetric backgrounds and by admitting a Birkhoff theorem which ensures that the corresponding spherically symmetric solutions are static and unique. This recent development has attracted significant attention from the community, as it provides a comprehensive and consistent framework for addressing the problem of regular black holes in a purely gravitational theory with a well-defined Einstein limit. It has allowed for the study of matter collapse leading to the formation of RBHs~\cite{Bueno:2024zsx,Bueno:2024eig,Bueno:2025gjg,Borissova:2026prd}, the stability of inner horizons~\cite{DiFilippo:2024mwm,Frolov:2026rcm,Bueno:2026wjt}, the interplay between the spacetime geometry and singular matter~\cite{Bueno:2025tli,Bueno:2026dln,Bueno:2026oyg} and others~\cite{Frolov:2024hhe,PinedoSoto:2026hfm,Konoplya:2024kih,Frolov:2026tft,Borissova:2026rbi,Konoplya:2024hfg,Ma:2024olw,Cisterna:2024ksz,Ditta:2024iky,Wang:2024zlq,Hennigar:2025yqm,Fernandes:2025eoc,Fernandes:2025fnz,Cisterna:2025vxk,Ling:2025ncw,Eichhorn:2025pgy,Fernandes:2025mic,Aguayo:2025xfi,Boyanov:2025pes,Konoplya:2025uta,Hao:2025utc,Tan:2025hht,Frolov:2025ddw,Chen:2025iuy,Arbelaez:2026eaz,Frolov:2026byy,Li:2026mam,Tsuda:2026xjc,Konoplya:2026gim,Fan:2026mdu,Bueno:2026bby}. Four-dimensional extensions are also possible, as long as one allows for non-polynomial densities in the gravitational action~\cite{Bueno:2025zaj,Borissova:2026wmn,Colleaux:2019ckh,Colleaux:2017ibe,Colleaux:2026hat}.

Most studies so far have naturally focused on establishing the existence of regular solutions, constructing explicit examples, and understanding their formation and physical properties. Comparatively less attention has been devoted to the global structure of the \textit{regular sector} itself. Once regularity can arise for broad families of higher-curvature couplings, several general questions become natural. What properties are shared by all regular vacuum solutions, independently of the particular resummed theory? Which types of regular cores are compatible with the vacuum equations? What constraints do the spherically symmetric vacuum spacetimes naturally satisfy? In particular, is regularity stable under continuous or sufficiently small deformations of the theory, or does it require selecting isolated points in theory space?

The purpose of this work is to address these questions for spherically symmetric vacuum QT solutions in a largely model-independent way. Rather than fixing a particular resummation of the higher-curvature series, we formulate the problem directly in terms of the characteristic function $h(\psi)$, introduced in Section~\ref{sec:qtg}, and derive consequences of the general conditions required for regularity.

The paper is organized as follows. In Section~\ref{sec:qtg} we review the relevant properties of QT theories and their spherically symmetric vacuum solutions. In Section~\ref{sec:degenerate} we discuss the degenerate branch of vacuum solutions, where the usual uniqueness properties are lost and more general geometries arise, such as redshifted Einstein static Universes and genuinely time-dependent vacua (see Appendix~\ref{app:degenerate}) while traversable wormholes remain excluded. In Section~\ref{sec:universal_core} we study the possible regular cores and show that every positive-mass regular vacuum solution necessarily possesses a de Sitter core. Section~\ref{sec:horizon_structure} is devoted to the horizon structure of these solutions,
as well as the possibility of multiple horizons and extremal points.  Finally, in Section~\ref{sec:space} we characterize the space of regular characteristic functions, study its topological properties, 
and analyze the stability of regularity under deformations of the higher-curvature couplings. Remarkably, we find that such a space of theories resembles the construction by Spanier on \textit{quasi-topological spaces}~\cite{Spanier:1963}, although the origin of the name is entirely unrelated to these particular theories. We will discuss such a connection, and conclude in Section~\ref{sec:conclusions}.

\section{Quasi-topological Gravity}\label{sec:qtg}
Consider a generic higher-order theory of gravity built from general contractions of the metric and the Riemann tensor, $\mathcal L(g^{ab},R_{abcd})$. The corresponding equations of motion can be written as~\cite{Padmanabhan:2011ex}
\begin{equation}
	\mathcal E_{ab} = P\indices{_a^{cde}}R_{bcde} - \frac{1}{2}g_{ab}\mathcal L - 2\nabla^c\nabla^dP_{acdb} = 8\pi GT_{ab}\,,
\end{equation}
with $P^{abcd} \equiv \partial \mathcal L /\partial R_{abcd}$. This indicates that, generically, higher-order theories of gravity have equations of motion featuring fourth-order derivatives of the metric.
Demanding second-order equations amounts to imposing the condition $\nabla^d P_{acdb} = 0$, which leads to Lovelock gravities
\cite{Lovelock:1970zsf,Lovelock:1971yv}.

One may relax this condition in various ways. In particular, consider a general spherically symmetric (SS) background
\begin{equation}
	\dd s^2_{\rm SS} = -N^2(t,r)f(t,r)\dd t^2 + \frac{\dd r^2}{f(t,r)} + r^2\dd\Omega^2_{D-2}\,.
	\label{eq:ds2SS}
\end{equation}
Demanding
$\nabla^d P_{acdb}\big|_{\rm SS} = 0$, namely, that the equations of motion become second order on such an ansatz,
	enlarges the space of theories to the so-called quasi-topological theories of gravity~\cite{Oliva:2010eb,Myers:2010ru,Dehghani:2011vu,Ahmed:2017jod,Cisterna:2017umf,Bueno:2019ycr,Bueno:2022res,Moreno:2023rfl,Bueno:2025qjk,Bueno:2026oyg,Murcia:2026pmg}. QTs---which include Lovelock theories as a subclass---are thus defined as the most general higher-order theories of gravity that admit second-order equations of motion on  SS backgrounds. As argued in~\cite{Bueno:2019ycr,Bueno:2022res,Moreno:2023rfl}, QTs exist at every curvature order $n \geq 2$ for every spacetime dimension $D\geq 5$ with the single exception of the $n=D/2$ case in even dimensions.\footnote{The fact that $n$-th order QT densities become trivial on spherically symmetric backgrounds in $D=2n$ (analogously to the Lovelock case) is the main historical reason behind the nomenclature ``quasi-topological'' (which was somewhat unfortunate prior to this paper).}

	In a general setup, let us consider the following action,
	\begin{equation}
		S_{\rm QT} = \frac{1}{16\pi G}\int\dd^Dx\sqrt{-g}\left(-2\Lambda + R + \sum_{n=2}^{N_{\rm max}}a_n \mathcal Z_n\right)\,,
		\label{eq:S_polyQTG}
	\end{equation}
	where $a_n$ are arbitrary couplings and $\mathcal Z_n$ are QT densities of curvature order $n$. In Appendix \ref{app:qt_densities}, one can consult instances of QT densities of curvature orders $n=1,\dots,5$, as well as a recursion formula which allows one to build densities of arbitrary curvature order using those as seeds.
	Additionally, it can be shown~\cite{Bueno:2024eig,Bueno:2024zsx} that the two-dimensional reduction of these theories on a $(D-2)$-sphere corresponds to a certain subclass of Horndeski theories~\cite{Horndeski:1974wa}.

	For the particular ansatz~\eqref{eq:ds2SS}, the full equations of motion of a general QT theory in the vacuum reduce to
	\begin{equation}
		h'(\psi)\partial_t f = 0\,,\; h'(\psi)\partial_r N = 0\,,\;\partial_r\left(r^{D-1}h(\psi)\right) = 0\,,	
		\label{eq:qtgSS_eqs}
	\end{equation}
	where 
	\begin{equation}
		\psi \equiv \frac{1-f(t,r)}{r^2}\,,
		\label{eq:psi}
	\end{equation}
	and where
	$h(\psi)$ is the so-called \textit{characteristic function}, defined as a power series in $\psi$ in the following way,\footnote{Naturally, an analogous characteristic polynomial controls the dynamics of SS Lovelock black holes, as previously pointed out \textit{e.g.,} in~\cite{Camanho:2011rj,Camanho:2013pzg}.}
	\begin{equation}
		h(\psi)\equiv -\frac{2\Lambda}{(D-1)(D-2)}+ \psi + \sum_{n=2}^{N_{\rm max}}\frac{D-2n}{D-2}a_n\psi^n\,.
		\label{eq:def_h}
	\end{equation}
	Hence, we are left with a set of first-order differential equations that can be directly integrated. Note that the only difference with respect to GR is fully encoded in the non-linearity of $h(\psi)$. Also, observe that in even dimensions, the term $\psi^{D/2}$ is absent for any choice of couplings. Solving the equations~\eqref{eq:qtgSS_eqs}, provided $h'(\psi)\neq 0$, we obtain
	\begin{equation}
		\partial_t f = 0\,,\quad \partial_r N = 0\,,\quad h(\psi) = \frac{2\mathsf M}{r^{D-1}}\,,
		\label{eq:qtgSS_eqs_2}
	\end{equation}
	for some integration constant $\mathsf M$ that is related to the ADM mass~\cite{Arnowitt:1960es,Arnowitt:1962hi} of the solution $M$ via
	\begin{equation}
		\mathsf M \equiv \frac{8\pi G M}{(D-2)\Omega_{D-2}}\,.
	\end{equation} 
	Throughout this work, unless otherwise specified, we will assume this quantity to be nonnegative, $\mathsf M\geq0$. The solutions to this set of equations correspond to $f(r,t) = f(r)$, and $N(r,t)=N(t)$, which can be set to unity $N(t)=1$ via a suitable time redefinition. Therefore, the solution is static, and it satisfies a Birkhoff theorem. The metric for such static and spherically symmetric (SSS) 	solutions then reads
	\begin{equation}
		\dd s^2 = -f(r)\dd t^2 + \frac{\dd r^2}{f(r)} + r^2\dd\Omega_{D-2}^2\,,
		\label{eq:ds2_SSS_N1}
	\end{equation}
	which satisfies $g_{tt}g_{rr}=-1$, and where the metric function can be written formally as
	\begin{equation}
		f(r) = 1-r^2\psi(r) = 1-r^2h^{-1}\left(\frac{2\mathsf M}{r^{D-1}}\right)\,,
	\end{equation}
	where $h^{-1}$ is the inverse of the characteristic function.
	
	Such an inverse can be ensured to exist whenever the characteristic function defining the theory is continuous and strictly monotonic on its domain. Since having a well-defined Einstein limit requires $h'(0)=1$, we require $h'(\psi)>0$ throughout the physical domain. This condition ensures the global existence of an inverse on the image of this domain. Additionally, the same formula would also apply, branch by branch, to \textit{e.g.} finite-order truncations of $h(\psi)$, which may admit several different real inverse branches.
	
	\subsection{Conditions for regularity}
	\label{subsec:condreg}
	As pointed out in~\cite{Bueno:2024dgm}, the SS solutions of broad classes of QTs containing an infinite tower of higher-order corrections are regular, with bounded curvature invariants. A set of sufficient conditions ensuring regularity was identified in~\cite{Bueno:2024dgm,Fernandes:2025fnz}. However, these conditions do not provide a complete characterization of the regular sector. A more general analysis is needed to identify the conditions that are both necessary and sufficient for all curvature invariants to remain bounded. To this end, let us introduce the Riemann tensor for the time-independent case of the spherically symmetric metric~\eqref{eq:ds2SS}, which can be written as~\cite{Deser:2005pc,Bueno:2025qjk,Bueno:2026dln}
	\begin{equation}
		\begin{aligned}
			R\indices{_{ab}^{cd}} &= \mathcal R^{(1)}\tau^{[c}_{[a}\rho^{d]}_{b]} + \mathcal R^{(2)}\tau^{[c}_{[a}\sigma^{d]}_{b]} \\
			&+ \mathcal R^{(3)}\rho^{[c}_{[a}\sigma^{d]}_{b]} + \mathcal R^{(4)}\sigma^{[c}_{[a}\sigma^{d]}_{b]}\,,
			\label{eq:general_Riem}
		\end{aligned}
	\end{equation}
	and where we have defined
	\begin{equation}
		\begin{aligned}
			\mathcal R^{(1)} &= -\frac{6f'N'+2Nf''+4fN''}{N}\,, \\
			\mathcal R^{(2)} &= -\frac{2Nf' + 4fN'}{rN}\,, \\
			\mathcal R^{(3)} &= -\frac{2f'}{r}\,,\quad
			\mathcal R^{(4)} = 2\frac{1-f}{r^2} = 2\psi\,.
		\end{aligned}
	\end{equation}
	Here, $\tau^b_a = \delta^t_a \delta^b_t$, $\rho_a^b = \delta^r_a \delta^b_r$ and $\sigma^b_a = \delta^b_a-\rho^b_a-\tau^b_a$ are projectors onto the orthogonal $t$, $r$ and angular directions, normalized such that $\tau_a^a=\rho_a^a=1$ and $\sigma_a^a=D-2$. Any algebraic curvature invariant of order $n$, obtained by contracting $n$ Riemann tensors with the metric, is a homogeneous polynomial of degree $n$ in the $\mathcal R^{(i)}$. Hence, boundedness of all four $\mathcal R^{(i)}$ is sufficient for all polynomial curvature invariants to remain bounded. It is also necessary: the quadratic invariant $R_{abcd}R^{abcd}$ is a positive sum of squares of the independent curvature components $\mathcal R^{(i)}$, and therefore its boundedness requires each $\mathcal R^{(i)}$ to be bounded. Thus, boundedness of the four functions $\mathcal R^{(i)}$ is necessary and sufficient for bounded polynomial curvature invariants.
	
	For the solutions of QT theories satisfying $h'(\psi)\neq 0$, each of the components $\mathcal R^{(i)}$ reduces to
	\begin{equation}
		\begin{aligned}
			\mathcal R^{(1)} &= 2r^2\psi'' + 8r\psi' + 4\psi\,, \\
			\mathcal R^{(2)} &= \mathcal R^{(3)} = 2r\psi' + 4\psi\,, \\
			\mathcal R^{(4)} &= 2\psi\,.
		\end{aligned}
	\end{equation}
	Therefore, boundedness of each $\mathcal R^{(i)}$ is equivalent to boundedness of $\psi$, $r\psi'$ and $r^2\psi''$. In what follows, we additionally restrict to solutions for which the curvature functions $\mathcal R^{(i)}$ admit finite limits. Under this assumption, in the particular $r\to 0$ limit, $\psi\to\psi_0$ must approach a constant, which further constrains $r\psi'$ and $r^2\psi''$ to vanish in this limit,\footnote{
		The case of $r\psi'$ converging to a non-vanishing constant in the $r\to 0$ limit would force $\psi\sim\log(r)$, contradicting bounded $\psi$, and yielding divergent curvature.
	} and hence implying the following necessary conditions in terms of the variable $\psi$ introduced in~\eqref{eq:psi},
	\begin{equation}
		\lim_{r\to 0}\psi(r) = \psi_0\,,\quad \lim_{r\to 0}r\psi'(r) = \lim_{r\to 0}r^2\psi''(r) = 0\,.
		\label{eq:conds_psi}
	\end{equation}
	Such conditions can be translated into requirements on the characteristic function $h(\psi)$ specifying the underlying theory. The field equation~\eqref{eq:qtgSS_eqs_2} implies $h(\psi)\propto r^{-(D-1)}$, so that $h(\psi)\to+\infty$ as $r\to0$. Moreover, the inversion between $\psi$ and $h(\psi)$ requires $h$ to be strictly monotonic. The existence of an Einstein gravity limit with positive Newton constant further requires $h'(0)=1$, and hence $h(\psi)$ must be monotone increasing throughout its domain. It is then straightforward to check that conditions~\eqref{eq:conds_psi} imply the following,
	\begin{equation}
		\lim_{\psi\to\psi_0^-}h(\psi) = +\infty\,,\quad \lim_{\psi\to\psi_0^-}\frac{h}{h'} = \lim_{\psi\to\psi_0^-}\frac{h^2h''}{(h')^3} = 0\,.
		\label{eq:conds_h}
	\end{equation}
	Therefore, any QT theory whose characteristic function is strictly monotonically increasing throughout its domain and satisfies the requirements~\eqref{eq:conds_h} admits vacuum spherically symmetric solutions with bounded curvature invariants. These conditions provide a direct characterization of regularity at the level of the characteristic function. 
	
	Their translation into necessary and sufficient constraints on the individual higher-order couplings, however, is considerably less transparent, and we will not attempt to pursue it here. Instead, it is useful to illustrate the above conditions with a simple explicit example.
	
	\subsubsection{Example: Hayward spacetime}
	A prototypical example of a spherically symmetric regular solution arising within this framework, in odd dimensions,\footnote{
		The restriction is due to the vanishing contribution of the curvature invariant of order $n=D/2$ on the spherical sector, for even dimensions.
	} is Hayward spacetime~\cite{Hayward:2005gi}. Indeed, setting the cosmological constant to zero, $\Lambda=0$, and choosing the higher-order couplings as 
	\begin{equation}
		a_n^{\rm H} = \frac{D-2}{D-2n}\alpha^{n-1}\,,
	\end{equation}
	where $\alpha$ has units of length$^2$, the corresponding infinite tower of higher-curvature corrections can be resummed into the characteristic function
	\begin{equation}
		h_{\rm H}(\psi)
		=
		\frac{\psi}{1-\alpha\psi}
		=
		\psi+\alpha\psi^2+\alpha^2\psi^3+\hdots
		\label{eq:h_Hayward}
	\end{equation}
	It can be shown that such a characteristic function satisfies the regularity conditions~\eqref{eq:conds_h}, with $\psi_0=1/\alpha$, which implies that close to $r=0$ the metric function $f_{\rm H}(r)$ will behave as
	\begin{equation}
		f_{\rm H}(r) = 1-\frac{r^2}{\alpha}+\hdots
	\end{equation}
	approaching a de Sitter-like core, and all curvature invariants will remain finite. The full solution follows from the inversion of the field equation~\eqref{eq:qtgSS_eqs_2}, yielding
	\begin{equation}
		f_{\rm H}(r) = 1-r^2\frac{2\mathsf M}{r^{D-1}+2\mathsf M\alpha}\,.
		\label{eq:f_Hayward}
	\end{equation}
	This corresponds to the higher-dimensional Hayward regular black hole, which should be regarded as a particularly simple representative of a much broader class of regular black hole geometries arising in these theories~\cite{Bueno:2024dgm,DiFilippo:2024mwm,Konoplya:2024hfg,Konoplya:2024kih,Frolov:2024hhe,Frolov:2026rcm,Bueno:2026dln,Tsuda:2026xjc}.
	
	\subsection{Effective energy conditions}\label{subsec:effective}
	Regular black holes, being singularity-free, must violate at least one of the assumptions of the singularity theorems, \textit{e.g.} the energy conditions~\cite{Borissova:2025msp,Borissova:2025hmj,Borissova:2026ohj}. In our setup, even though we are in the vacuum, we can still adopt an effective point of view and regard the higher-curvature contributions to the field equations as an effective matter content in Einstein gravity, writing $G_{ab}=8\pi G T_{ab}^{\rm eff}$, together with $\nabla^a T_{ab}^{\rm eff}=0$. From now on, we shall refer to the energy conditions as \textit{effective} energy conditions, or curvature conditions, since they will be imposed at the level of the curvature rather than the matter content itself. 
	
	Some violation of the effective strong energy condition (SEC) is expected to happen as one approaches the center of the black hole \cite{Zaslavskii:2010qz}, reflecting the ``repulsive'' behavior that helps resolve the spacetime singularity. This is not necessarily troublesome, as a positive cosmological constant also violates the SEC. A violation of the null energy condition (NEC), on the other hand, would be more concerning. The NEC requires that $T_{ab}^{\rm eff}k^ak^b\geq0$ for all future-directed null vectors $k^a$, or equivalently $G_{ab}k^ak^b\geq0$. In this section, we will review the main energy conditions \cite{Maeda:2018hqu}, and translate them into constraints imposed directly on $h(\psi)$ defining the underlying theory.
	
	\begin{table*}[t]
		\centering
		\renewcommand{\arraystretch}{1.8}
		\setlength{\tabcolsep}{10pt}
		\renewcommand{\arraystretch}{1.7}
		\begin{tabular}{c|c|l}
			(\textit{Effective}) Energy condition & Curvature condition & \multicolumn{1}{c}{Constraint on $h$} \\ \hline\hline
			(Null) NEC & $G_{ab}k^ak^b\geq0$ & $h''\geq0$ \\
			(Weak) WEC & $G_{ab}v^av^b\geq0$ & $h''\geq0\,,\quad \psi h'-h\geq0$ \\
			(Strong) SEC & $R_{ab}v^av^b\geq0$ & $h''\geq0\,,\quad (D-1)h^2h''\geq2(h')^2(\psi h'-h)$ \\
			(Dominant) DEC & $G_{ab}v^av^b\geq0\,,\quad J_aJ^a\leq0$ & $0\leq(D-1)h^2h''\leq2(D-2)(h')^2(\psi h'-h)$
		\end{tabular}
		\caption{Effective energy conditions for arbitrary static and spherically symmetric vacuum solutions of the form \eqref{eq:ds2_SSS_N1}, on a branch with $h'>0$. The curvature inequalities must hold for every future-directed null vector $k^a$ or unit timelike vector $v^a$, as appropriate, and $J^a\equiv-G^a{}_bv^b$. Primes denote derivatives with respect to $\psi$, and all constraints on $h$ are imposed throughout the physical domain. The relations between the various conditions are, schematically, ${\rm DEC}\Rightarrow \rm{WEC}\Rightarrow \rm{NEC}\Leftarrow \rm{SEC}$.}
		\label{tab:effective_energy_conditions}
	\end{table*}
	
	Table~\ref{tab:effective_energy_conditions} summarizes the main results, relating every effective energy condition to inequality bounds that both the characteristic function $h$ and its derivatives must satisfy. In particular, on a branch with $h'>0$, the effective NEC implies $h''\geq0$, \textit{i.e.}, the characteristic function must be convex. Notice that for asymptotically flat solutions with $h(0)=0$ and $\psi\geq0$, this also ensures the WEC, since $\psi h'-h=\int_0^\psi u h''(u)\dd u\geq0$. On the other hand, the SEC imposes an additional inequality that must necessarily be violated somewhere in the interior of a regular black hole. The dominant energy condition (DEC) imposes further constraints that do not follow from convexity alone.
	
	\begin{figure}[t]
		\centering
		\includegraphics[width=\linewidth]{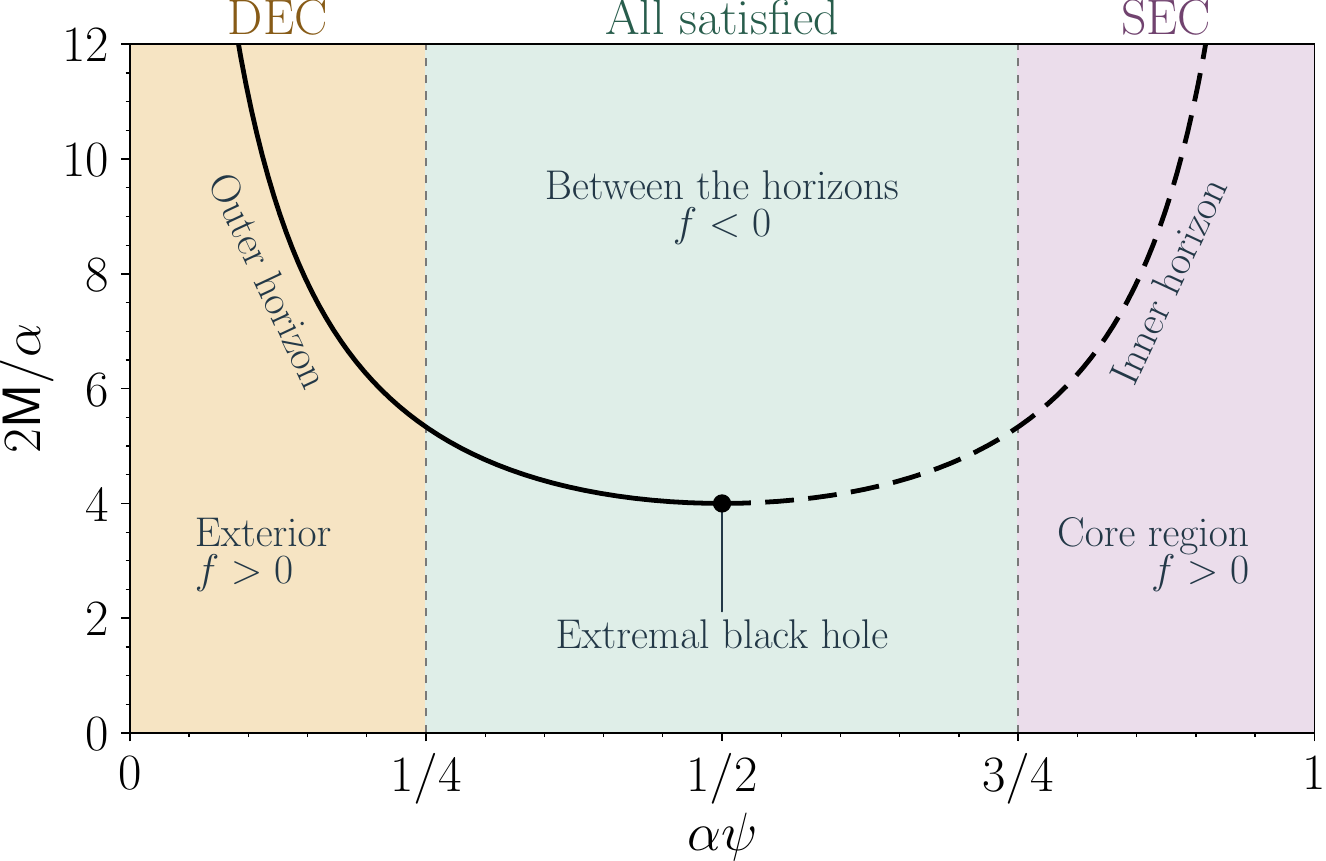}
		\caption{Regions of violation of the effective energy conditions for the Hayward solution \eqref{eq:f_Hayward} in $D=5$ at fixed $\alpha>0$. Shaded regions indicate violation of the DEC (left), sufficiently far away from the horizon, whereas violation of the SEC happens in the interior of the black hole. The green shaded region indicates that all conditions are satisfied there. The dark curve locates the horizons, obtained from inversion of \eqref{eq:qtgSS_eqs_2} upon substituting $\psi_h = 1/r_h^2$. Its two branches merge at the extremal point, located at $(\alpha\psi_{\rm ex},2\mathsf M_{\rm ex}/\alpha)=(1/2,4)$.}
		\label{fig:energy_conds_HaywardD5}
	\end{figure}
	
	For the Hayward solution \eqref{eq:f_Hayward}, these features are illustrated in Fig.~\ref{fig:energy_conds_HaywardD5}, where the mass is plotted against $\psi$, for fixed $\alpha>0$ and $D=5$. Each horizontal slice corresponds to a fixed-mass solution, with $\psi$ increasing from spatial infinity toward the regular center. The NEC and WEC hold everywhere. Additionally, the SEC is satisfied whenever
	\begin{equation}
		\psi\leq \frac{D-2}{D-1}\frac{1}{\alpha}\,,
	\end{equation}
	which reduces to $\psi\leq 3/(4\alpha)$ in $D=5$, as shown in Fig.~\ref{fig:energy_conds_HaywardD5}. On the other hand, the DEC is satisfied whenever
	\begin{equation}
		\psi \geq\frac{1}{D-1}\frac{1}{\alpha}\,,
	\end{equation}
	which reduces to the inequality bound $\psi\geq 1/(4\alpha)$ for $D=5$ that is shown in Fig.~\ref{fig:energy_conds_HaywardD5}. Therefore, all effective energy conditions are simultaneously satisfied for $1/4\leq \alpha\psi \leq 3/4$. In particular, for sufficiently near-extremal black holes, both horizons and the entire region between them lie within this interval. Thus, violations of the effective energy conditions can be confined away from the horizons, occurring near the regular core for the SEC and in the asymptotic region for the DEC. The violation of the DEC asymptotically far away from the outer horizon can be understood as follows: 
	the effective stress-energy tensor admits the expansion $(T_a^b)^{\rm eff} = {\rm diag}(-\rho,p_r,p_t,\hdots,p_t)$, where $\rho$, $p_r$ and $p_t$ are the effective energy density, radial pressure and the $(D-2)$ tangential pressures. For such tensors, the DEC requires that $\rho^2\geq p_i^2$, for every principal pressure $p_i$. This guarantees that the energy-momentum flux measured by any timelike observer is causal. Since $p_r=-\rho$ for spacetimes of the form \eqref{eq:ds2_SSS_N1}, the radial condition is always saturated. For the tangential condition, however, asymptotically far away from the horizon we have the power-law decay $p_t(r)\sim D\rho(r)/(D-2)\sim \mathsf M^2\alpha/r^{2(D-1)}$. Even though the effective energy-momentum tensor becomes arbitrarily small as $r\to\infty$, the DEC remains violated at every sufficiently large finite radius \cite{Lan:2022bld}.
	
	These results illustrate how the characteristic function constrains the physical properties of the corresponding vacuum solutions. 
	
	\section{Degenerate branch of solutions}\label{sec:degenerate}
	
	It is interesting to note that equations~\eqref{eq:qtgSS_eqs}, describing the spherically symmetric solutions of generic QT theories, are also satisfied if $h(\psi)=h'(\psi) = 0$. In this situation, in principle, the pieces $\partial_r N$ and $\partial_t f$ are not constrained to vanish on-shell, leading to a degeneracy in the space of solutions \cite{Bueno:2025qjk}. 
	
	Since $h$ is analytic and non-constant (in particular, $h'(0)=1$), these conditions can only be satisfied at isolated values of $\psi$, which forces $\psi(r,t)$ to be constant. A simple example is the following,
	\begin{equation}
		h(\psi) = -\frac{2\Lambda}{(D-1)(D-2)} + \psi + \frac{D-4}{D-2}a_2\psi^2\,,
	\end{equation}
	representative of Einstein-Gauss-Bonnet (EGB) gravity with a cosmological constant term.
	
	It is possible to find a relation between the cosmological constant $\Lambda$ and the quadratic coupling $a_2$ so as to satisfy the degeneracy conditions at a certain value $\psi=\psi_0(\Lambda,a_2)$. Indeed, it is straightforward to see this is the case for\footnote{
		This particular relation among the couplings of the theory is known as the Chern-Simons point of EGB~\cite{Anabalon:2024abz}, as it is possible to reformulate the theory in five dimensions and with a negative cosmological constant into an $SO(4,2)$ gauge theory featuring a Chern-Simons term in the action~\cite{Chamseddine:1989nu}. This point in parameter space admits additional features (see~\cite{Anabalon:2009kq,Anabalon:2010ns,Camanho:2014apa,Dotti:2006cp,Dotti:2007az,Giribet:2014hpa} and references therein).
	}
	\begin{equation}
		\psi_0 = -\frac{1}{2a_2}\frac{D-2}{D-4}\,,\quad \Lambda = -\frac{(D-1)(D-2)^2}{8(D-4)}\frac{1}{a_2}.
		\label{eq:conds_egb_ft}
	\end{equation}
	Note that $\psi_0>0$ requires $a_2<0$, and hence $\Lambda>0$.
	Since $\psi(r,t)=\psi_0$ is constrained to be constant, then $f(r,t)=f(r)=1-\psi_0 r^2$, while $N(r,t)$ is arbitrary. Therefore, spacetimes of the form
	\begin{equation}
		\dd s^2_{\rm deg} = -A(r,t)\dd t^2 + \frac{\dd r^2}{1-\psi_0 r^2} + r^2\dd\Omega_{D-2}^2\,,
		\label{eq:ds2_deg}
	\end{equation}
	where we have defined $A(r,t)\equiv N^2(r,t)(1-\psi_0 r^2)$,
	will be solutions of QT theories~\eqref{eq:S_polyQTG} for any arbitrary $N(r,t)$, as long as $\psi_0$ satisfies the conditions $h(\psi_0)=h'(\psi_0)=0$. Since the $g_{tt}$ component remains arbitrary, this degenerate branch contains a broad family of geometrically distinct spacetimes (see Appendix~\ref{app:degenerate} for some representative examples).
	
	The freedom is broad, and one can in principle engineer a wide variety of geometries, including: Einstein's static Universe, static redshifted deformations, and even time-dependent vacuum configurations. Regular geometries can also be obtained, provided \(A(r,t)\) satisfies the corresponding regularity conditions.
	
	The drawback, however, is that these solutions are fine-tuned: they exist only on a special codimension-one locus in the space of couplings, defined by the degeneracy conditions, such as~\eqref{eq:conds_egb_ft} in the EGB example, but in general $h(\psi_0)=h'(\psi_0)=0$ for some $\psi_0$.
	
	In the following section, we will study in more detail the structure of the regular sector of vacuum solutions of generic QTs and show that, beyond these fine-tuned configurations, regular solutions persist over a much broader region of theory space.

	\section{Universality of the core}\label{sec:universal_core}
	
	The notion of regularity adopted throughout this paper is based on the boundedness of polynomial curvature invariants. 
	In particular, notice that the regularity conditions \eqref{eq:conds_psi} imply that the Riemann tensor admits the following expression in the $r=0$ limit,
	\begin{equation}
		\begin{aligned}
			R\indices{_{ab}^{cd}}&\to 2\psi_0\left(2\tau^{[c}_{[a}\rho^{d]}_{b]} + 2\tau^{[c}_{[a}\sigma^{d]}_{b]}+2\rho^{[c}_{[a}\sigma^{d]}_{b]}+\sigma^{[c}_{[a}\sigma^{d]}_{b]}\right) \\
			&= \psi_0\left(\delta_a^c \delta_b^d-\delta_a^d\delta_b^c\right)\,,
		\end{aligned}
	\end{equation}
	where we have used the properties of the projectors, together with $\tau_a^b+\rho_a^b+\sigma_a^b = \delta_a^b$. Hence, regular spacetimes approach a locally maximally symmetric region where the curvature approaches the value $R= D(D-1)\psi_0$. We therefore classify the core according to the sign of the Ricci scalar: it is \textit{de Sitter} (dS) if $R>0$, \textit{anti-de Sitter} (AdS) if $R<0$, and Minkowski if $R=0$, the sign of $R$ being that of $\psi_0$.
	
	Within the class of QTs admitting regular black hole solutions, we will prove that every positive-mass regular vacuum solution necessarily possesses a positively curved de Sitter core. Consequently, geometries with anti-de Sitter or Minkowski cores cannot arise as non-trivial regular vacuum solutions of these theories.
	
	As shown in Sec.~\ref{sec:qtg}, if we consider monotonic characteristic functions $h'(\psi)\neq0$, the solutions are static, satisfying a Birkhoff theorem, and solely described by a single metric function given by $f(r)=1-r^2\psi(r)$, which can be obtained by inverting the algebraic equation~\eqref{eq:qtgSS_eqs_2}. Following the notation of~\cite{Hennigar:2025yqm}, let us define the \textit{asymptotically flat} part of the characteristic function as 
	\begin{equation}
		h_{\rm f}(\psi) \equiv \psi + \sum_{n=2}^{N_{\rm max}}\frac{D-2n}{D-2}a_n\psi^n\,,
	\end{equation}
	which corresponds to the characteristic function of asymptotically flat spacetimes, with vanishing cosmological constant $\Lambda=0$.
	Let us also rewrite the couplings as 
	\begin{equation}
		\frac{D-2n}{D-2}a_n \equiv \alpha_n\,,
		\label{eq:an_alphan}
	\end{equation}
	and introduce the cosmological lengthscale via
	\begin{equation}
		\frac{2\Lambda}{(D-1)(D-2)}\equiv \frac{\sigma}{\ell^2}\,,
	\end{equation}
	where $\sigma\in\{+1,0,-1\}$ accounts for the sign of the cosmological constant term. In this case, the algebraic equation of motion of the theory can be rewritten as
	\begin{equation}
		h_{\rm f}(\psi) \equiv \psi + \sum_{n=2}^{N_{\rm max}}\alpha_n\psi^n = \frac{2\mathsf M}{r^{D-1}} + \frac{\sigma}{\ell^2} \equiv \mathcal S(r).
		\label{eq:hf_S}
	\end{equation}
	With these redefinitions, the metric function can be formally rewritten as
	\begin{equation}
		f(r) = 1-r^2\psi(r) = 1-r^2h_{\rm f}^{-1}[\mathcal S(r)]\,.
		\label{eq:f_general}
	\end{equation}
	For our purposes, we will work with theories including an infinite tower of higher-order corrections, $N_{\rm max}\to\infty$, as well as consider the characteristic function to diverge at a finite value of the argument, $h(\psi)\rightarrow \infty$ as $\psi \rightarrow \psi_0^-$.
	
	In full generality, the couplings of the theory can be written as $\alpha_n = \beta_n \alpha^{n-1}$, for some parameter $\alpha>0$ with units of length$^2$, specifying the characteristic length scale at which the higher-order effects become important, and where $\beta_n$ are theory-dependent dimensionless couplings.
	
	It is also important to note that generic higher-order theories of gravity modify the expressions for the gravitational coupling $G$ and the cosmological constant $\Lambda$, promoting them to effective parameters that depend on the higher-order couplings of the theory. The relation is fully encoded in the characteristic function, given by~\cite{Hennigar:2025yqm}
	\begin{equation}
		\ell_{\rm eff}^2 = \frac{\sigma}{h_{\rm f}^{-1}(\sigma/\ell^2)}\,,\quad G_{\rm eff} = \frac{G}{h_{\rm f}'(\sigma/\ell_{\rm eff}^2)}\,.
	\end{equation}
	Note that if $h_{\rm f}'(\psi)<0$ for some values of $\psi$, the effective gravitational coupling may become negative, which is physically undesirable and, in the present context, precludes regularity of the black hole solutions. This can be avoided by requiring $h_{\rm f}'(\psi)>0$ throughout the physical domain $r\in \mathbb R^+$,
	which also guarantees the existence of the inverse function $h_{\rm f}^{-1}(\psi)$, as previously mentioned. Furthermore, depending on the sign of $h_{\rm f}^{-1}(\sigma/\ell^2)$, the effective cosmological constant may have the opposite sign to the bare one, corresponding to $\ell_{\rm eff}^2<0$. We will not consider this possibility, and assume that $h_{\rm f}^{-1}(\sigma/\ell^2)$ and $\sigma$ have the same sign, which is the physically interesting scenario,\footnote{This assumption is not merely technical. For instance, for $h_{\rm f}=\psi/(1+\alpha\psi)$ with $\sigma=+1$, $\ell^2<\alpha$ and $\mathsf M>0$, one finds $\psi=\mathcal S/(1-\alpha\mathcal S)<-1/\alpha$ for all $r$, which describes a regular, horizonless solution with an AdS core and AdS asymptotics, living on a branch disconnected from $\psi=0$.}
	as these solutions will correspond to the Einstein branch.

	In this situation, provided the function $\mathcal S(r)$ appearing in the equations of motion is monotonic,
	\begin{equation}
		\mathcal S'(r) = -\frac{(D-1)2\mathsf M}{r^{D}} < 0\quad\forall r\in\mathbb R^+\,,
	\end{equation}
	which happens if the mass $\mathsf M>0$,\footnote{For $\mathsf M<0$, $\mathcal S(r)\to-\infty$ as $r\to0$ and $\psi(r)$ increases with $r$. A regular core then requires $h_{\rm f}\to-\infty$ at a finite negative value of $\psi$, and the core is AdS. Hence, ${\rm sign}(\psi_0)={\rm sign}(\mathsf M)$.}
	it follows that
	\begin{equation}
		\psi'(r) = \partial_r h_{\rm f}^{-1}[\mathcal S(r)] = -\frac{(D-1)2\mathsf M}{r^Dh_{\rm f}'(\psi)} < 0\,,
	\end{equation}
	since $h_{\rm f}'(\psi)>0$. Therefore, the function
	\begin{equation}
		\psi(r) = \frac{1-f(r)}{r^2}
	\end{equation}
	is monotonically decreasing throughout the physical domain $r>0$.
	
	This monotonicity imposes strong constraints on the possible vacuum geometries. In particular, it relates the value of $\psi$ at the center,
	\begin{equation}
		\psi_0 \equiv \lim_{r\to 0}\psi(r)\,,
	\end{equation}
	to its asymptotic value,
	\begin{equation}
		\psi_\infty \equiv \lim_{r\to\infty}\psi(r) = \frac{\sigma}{\ell_{\rm eff}^2}<\psi_0\,.
	\end{equation}
	As discussed in Sec.~\ref{sec:qtg}, the near-origin expansion of the metric reads $f(r) \sim 1-\psi_0 r^2$. Consequently, $\psi_0>0$ corresponds to a de Sitter (dS) core, $\psi_0<0$ to an anti-de Sitter core, and $\psi_0=0$ to a Minkowski core.
	
	\subsubsection{de Sitter core}
	For a de Sitter core, $\psi_0$ must be positive. Since $\psi(r)$ decreases monotonically, one necessarily has
	\begin{equation}
		\psi_\infty < \psi_0\,.
		\label{eq:ineq_psi}
	\end{equation}
	Since $\psi_\infty = \sigma/\ell_{\rm eff}^2$ and (re)defining $\alpha \equiv 1/\psi_0$,
	with units of $[\alpha]=L^2$, as the scale at which the higher-order terms in the action become important,
	we obtain
	\begin{equation}
		\frac{\sigma}{\ell_{\rm eff}^2} < \frac{1}{\alpha}\,.
	\end{equation}
	This inequality is trivially satisfied for asymptotically AdS ($\sigma=-1$) and asymptotically flat $(\sigma=0)$ spacetimes. For asymptotically de Sitter solutions $(\sigma=+1)$, however, it translates into the bound
	$\ell_{\rm eff}^2 > \alpha$, which is automatically satisfied: the curvature of the asymptotic dS vacuum is always smaller than that of the core.
	
	Constructing theories whose regular and spherically symmetric vacuum solutions preserve a regular dS core is straightforward (\textit{e.g.} Hayward), as it suffices to choose an $h(\psi)$ that diverges at a finite value $\psi=\psi_0>0$, and satisfies~\eqref{eq:conds_h}.
	
	\subsubsection{Minkowski core}
	For a Minkowski core, $\psi_0$ vanishes. In this case, the monotonicity of $\psi(r)$ implies $\psi_\infty \leq 0$. The limiting case $\psi_\infty = 0$, which corresponds to asymptotically flat spacetimes, requires $\psi(r)\equiv 0$ for all $r>0$, so that the spacetime is exactly Minkowski, corresponding to the unique vacuum solution with vanishing mass and cosmological constant, namely $\mathsf M = 0$ and $\Lambda = 0$.
	
	One may wonder whether the strict inequality
	\begin{equation}
		\psi_\infty=-\frac{1}{\ell_{\rm eff}^2}<0=\psi_0
	\end{equation}
	could instead describe a vacuum spacetime with a Minkowski core that is asymptotically AdS. This possibility is excluded. Indeed, for $\mathsf M>0$, the equations of motion imply that $\mathcal S(r)\to\infty$ as $r\to0$, whereas $\psi(r)\to\psi_0=0$ implies that $h_{\rm f}(\psi)\to0$. Since the equations of motion require $h_{\rm f}(\psi)=\mathcal S(r)$, this leads to a contradiction, since the right-hand side diverges while the left-hand side vanishes. Hence, the only vacuum solution with a Minkowski core is the trivial Minkowski spacetime, corresponding to $\mathsf M=0$ and $\Lambda=0$.
	
	\subsubsection{Anti-de Sitter core}
	For an AdS core, $\psi_0\equiv -1/\alpha<0$. In this case, monotonicity requires the spacetime to be asymptotically AdS, with
	\begin{equation}
		\frac{1}{\ell_{\rm eff}^2}>\frac{1}{\alpha}\,,
	\end{equation}
	as otherwise inequality~\eqref{eq:ineq_psi} would be violated. Nevertheless, no positive-mass vacuum solution with an AdS core exists within the class of QTs under consideration. Indeed, 
	since
	$h_{\rm f}(\psi)$ is monotonically increasing and vanishes only at $\psi=0$, the condition $\psi(r)<\psi_0<0$ implies $h_{\rm f}(\psi(r))<0$ for all $r>0$, whereas $\mathcal{S}(r)\rightarrow +\infty$ as $r\rightarrow 0$,
	yielding a contradiction. Therefore, positive-mass regular vacuum solutions with AdS cores are excluded. The only remaining possibility is the massless solution, $\mathsf M=0$, for which $\psi(r)=\psi_0<0$ $\forall r\in\mathbb R^+$ is constant, corresponding to pure AdS spacetime.
	
	We have thus proven the following no-go result, and conclude that, within the class of QTs considered in the present work,\footnote{
		Note, however, that this proof relies crucially on the particular form of the algebraic equation~\eqref{eq:hf_S}. Certain non-polynomial theories of the  quasi-topological kind---in the sense that they admit an algebraic equation for the metric function $f(r)$, as well as admitting a Birkhoff theorem---, whose algebraic equation is not of that particular form may very well have this type of additional solutions in the vacuum~\cite{Borissova:2026krh,Mazza:2026taj,Colleaux:2026hat}.
	} every positive-mass regular vacuum solution necessarily possesses a positively curved (de Sitter) core. Minkowski and anti-de Sitter cores are realized only by the trivial maximally symmetric vacuum solutions. 
	
	Note that the vacuum assumption is essential to this result. In the presence of matter, regular black hole solutions with anti-de Sitter cores can exist---see, \textit{e.g.},~\cite{Hennigar:2025yqm}, where such cores arise in theories coupled to nonlinear electrodynamics (see also \cite{Arrechea:2025nlq} for a more phenomenological construction).
	
	\section{Horizon structure}\label{sec:horizon_structure}
	In the previous section we have studied the deep interior of the regular vacuum solutions of quasi-topological gravity, and concluded that the only non-trivial, positive-mass configurations are those whose core is de Sitter, and therefore positively curved. In this section we will analyze in a general way the horizon structure of such vacuum regular configurations.
	
	\subsection{Existence of extremal configurations}
	\subsubsection{Asymptotically flat case}
	For simplicity, we begin the discussion with those solutions that tend to the asymptotically flat  Schwarzschild-Tangherlini solution $f(r)\sim 1-2\mathsf M/r^{D-3}$ for values of $r$ much larger than the characteristic scale of higher-curvature terms.
	
	For spacetimes of the form~\eqref{eq:ds2_SSS_N1}, a Killing horizon satisfies $f(r_h)=0$, which implies that
	\begin{equation}
		\psi_h\equiv \psi(r_h) = \frac{1-f(r_h)}{r_h^2} = \frac{1}{r_h^2} > 0\,.
	\end{equation}
	Substituting this relation back into the algebraic equation~\eqref{eq:hf_S} we obtain the following on-shell expression,
	\begin{equation}
		\mathsf M(r_h) = \frac{r_h^{D-1}}{2}h_{\rm f}\left(\frac{1}{r_h^2}\right)\,.
		\label{eq:M_rh}
	\end{equation}
	Since the physical branch, corresponding to $r\in\mathbb R^+$, satisfies $0\leq \psi < \psi_0$, horizons can only occur for
	\begin{equation}
		r_h > r_0 \equiv \frac{1}{\sqrt{\psi_0}}\,.
	\end{equation}
	It is interesting to note that this lower bound determines the behavior of~\eqref{eq:M_rh} at both endpoints. When the horizon approaches $r_h\to r_0^+$, we have $\psi_h\to\psi_0^-$ and expression $\mathsf M(r_h)\propto h_{\rm f}(\psi_h)$ diverges. On the other hand, when $r_h\to+\infty$, we have $\mathsf M(r_h)\propto r_h^{D-3}$, which also diverges. Therefore, the continuous function $\mathsf M(r_h)$ diverges to positive infinity at both endpoints, which means that it must attain at least one minimum at a finite radius. 
	
	Differentiating~\eqref{eq:M_rh} with respect to the horizon location, we obtain
	\begin{equation}
		\dv{\mathsf M}{r_h} = \frac{r_h^{D-2}}{2}\left[(D-1)h_{\rm f}(\psi_h) - 2\psi_h h'_{\rm f}(\psi_h)\right]\,.
		\label{eq:dM_drh}
	\end{equation}
	Therefore, every minimum of $\mathsf M(r_h)$ will satisfy
	\begin{equation}
		(D-1)h_{\rm f}(\psi_m) - 2\psi_m h'_{\rm f}(\psi_m) = 0\,.
		\label{eq:extr_cond}
	\end{equation}
	This condition is precisely the  extremality condition $f(r_h)=f'(r_h)=0$. Indeed, from expression~\eqref{eq:f_general} for the metric function, we can write
	\begin{equation}
		f'(r_h) = r_h\left[(D-1)\frac{h_{\rm f}(\psi_h)}{h'_{\rm f}(\psi_h)}-2\psi_h\right]\,,
	\end{equation}
	which vanishes exactly when~\eqref{eq:extr_cond} is satisfied. We can rewrite that condition as
	\begin{equation}
		\dv{\log h_{\rm f}}{\log \psi}\bigg|_{\psi_m} = \frac{D-1}{2}>1\,.
		\label{eq:extr_cond_rewritten}
	\end{equation}
	
	In the weak-field limit, $h_{\rm f}(\psi)\sim\psi + \order{\psi^2}$, and therefore $\dd\log h_{\rm f}/\dd\log\psi = 1+\order{\psi}$. On the other hand, as $\psi\to\psi_0^-$, the left-hand side of~\eqref{eq:extr_cond_rewritten} diverges, since $h_{\rm f}(\psi)\to+\infty$ at a finite value of $\psi_0$. Hence, by continuity, there must exist at least one point $\psi_m\in(0,\psi_0)$ satisfying~\eqref{eq:extr_cond}, corresponding to an extremal black hole solution~\cite{Bueno:2025tli}.
	
	Notice, however, that there is in principle no upper bound on the number of solutions of~\eqref{eq:extr_cond_rewritten}. If the logarithmic slope of $h_{\rm f}$ is monotonic, \textit{i.e.} $(\dd\log h_{\rm f}/\dd\log\psi)'>0$, where prime denotes differentiation with respect to $\psi$, then the extremal solution is unique. In general, however, the monotonicity of this quantity depends on higher derivatives of $h_{\rm f}$, and in particular of $h''_{\rm f}(\psi)$, which is not required to have a definite sign, unlike $h'_{\rm f}(\psi)$. Therefore, unless additional restrictions are imposed on the behavior of $h_{\rm f}''(\psi)$, or more generally on the logarithmic slope itself, there is no upper bound on the number of extremal black hole solutions, and such solutions will, in general, correspond to different values of the mass.
	On the other hand, if all the couplings are nonnegative, $\alpha_n\geq 0$, and $\alpha_1=1$, the derivative of the logarithmic slope can be written as
	\begin{equation}
		\psi\dv{\psi}\left(\dv{\log h_{\rm f}}{\log \psi}\right) = \frac{1}{2h_{\rm f}^2}\sum_{n,m\geq 1}(n-m)^2\alpha_n\alpha_m\psi^{n+m}>0\,,
	\end{equation}
	which is positive definite, meaning the logarithmic slope $\dd \log h_{\rm f}/\dd\log \psi$ is monotonically increasing and therefore there is a unique solution to the extremality condition~\eqref{eq:extr_cond_rewritten}.

	\subsubsection{Asymptotically AdS spacetimes}
	The previous result can be easily generalized when considering a negative cosmological constant. From~\eqref{eq:hf_S}, and for $\sigma=-1$, the horizon equation becomes
	\begin{equation}
		\mathsf M(r_h) = \frac{r_h^{D-1}}{2}h\left(\frac{1}{r_h^2}\right)\,,
		\label{eq:M_rh_AdS}
	\end{equation}
	and the extremality condition $\dd\mathsf M/\dd r_h = 0$ will read
	\begin{equation}
		\dv{\log h}{\log \psi}\bigg|_{\psi_h} = \psi_h\frac{h'(\psi_h)}{h(\psi_h)} = \frac{D-1}{2}\,,
		\label{eq:extr_cond_rewritten_AdS}
	\end{equation}
	where now $h(\psi) = 1/\ell^2 + h_{\rm f}(\psi)$ is the full characteristic function. For asymptotically anti-de Sitter spacetimes, $f(r)\sim 1+r^2/\ell_{\rm eff}^2$ in the large-radius limit, which means that $\psi_\infty = -1/\ell_{\rm eff}^2<0$. 
	Horizons correspond again to $\psi_h=1/r_h^2\in(0,\psi_0)$. In the limit $\psi_h\to0^+$, \textit{i.e.} $r_h\to\infty$, one has $h\to1/\ell^2\neq0$ and hence $\dd\log h/\dd\log\psi\to0$, while it diverges in the $\psi_h\to\psi_0^-$ limit.  
	Therefore, by continuity,
	there must exist a point that satisfies~\eqref{eq:extr_cond_rewritten_AdS}, which is the extremal point. The same argument of the previous subsection applies, and we cannot put an upper bound on the number of extremal points as long as we do not impose additional restrictions on the behavior of $h''_{\rm f}(\psi)$.  If $\alpha_n\geq0$, however, $h$ is still a power series with nonnegative coefficients, and the extremal point is again unique.

	We therefore extend the result of the previous section to asymptotically anti-de Sitter spacetimes. The corresponding analysis for asymptotically de Sitter spacetimes is left for future work.
	
	\subsection{Unboundedness of the number of horizons}
	We have mentioned that, in general, as long as we do not impose further constraints on the monotonicity of the logarithmic slope of $h(\psi)$, then the number of solutions of the extremality condition and of the horizon condition $f(r_h)=0$ is not bounded from above. 
	This means that we can have, in principle, spherically symmetric black holes with multiple inner horizons or extremal black holes of different masses, among others.
	
	We can construct a family of theories satisfying this property by a suitable deformation of some known theory that yields regular black holes with the simplest horizon structure (at most two horizons), \textit{e.g.} Hayward.
	For instance, consider the theory defined by the following characteristic function,
	\begin{equation}
		h_N(\psi) = C\psi^2 + \frac{\psi}{1-\alpha\psi}P_N(\psi)\,,
		\label{eq:hN}
	\end{equation}
	where $P_N(\psi)$ is a polynomial of degree $2N$, defined by
	\begin{equation}
		P_N(\psi) \equiv \prod_{j=1}^N\left(1-\frac{\psi}{c_j}\right)^2\,,
	\end{equation}
	and with roots located at $\psi_j=c_j$. Here, $N\in\mathbb N$, as well as $C>0$, are parameters that we can choose freely. Note that such a theory has a well-defined Einstein limit, provided $h_N(0) = 0$ and $h_N'(0)=1$. At the same time, as $\psi\to\psi_0^-$, the characteristic function diverges as $h_N(\psi)\sim P_N(\psi_0)h_{\rm H}(\psi)$, with $h_{\rm H}(\psi)$ the theory describing the Hayward solution, given in \eqref{eq:h_Hayward}, implying that such a theory yields a regular solution. To further show that such a theory yields a multiple horizon black hole, observe that the horizon equation reads
	\begin{equation}
		2\mathsf M = \frac{h_N(\psi_h)}{\psi_h^2} = C + \frac{P_N(\psi_h)}{\psi_h(1-\alpha\psi_h)}\,,
	\end{equation}
	where we have considered $D=5$ for simplicity. The second term vanishes at each $\psi_j=c_j$, is strictly positive between those points, and diverges at both endpoints $\psi\to 0^+$ and $\psi\to\psi_0^-$. Consequently, the function $h_N(\psi)/\psi^2$ has $N$ distinct minima, all at height $C$. For $2\mathsf M = C+\epsilon$, with $\epsilon>0$ sufficiently small, there are two intersections near each minimum, which gives at least $2N$ distinct horizons at the same mass. Since $N$ is arbitrary, there is no upper bound on the number of horizons, and therefore we show a precise way of constructing theories whose unique spherically symmetric vacuum solutions contain multiple inner horizons.
	
	One could argue that such a construction must necessarily violate one (or some) of the effective energy conditions discussed in Section \ref{subsec:effective}. In particular the SEC, which, as expected, is violated in the vicinity of the de Sitter core, is also violated locally near the several inner horizons \cite{Yang:2021civ}. This is not necessarily problematic. A violation of the NEC, on the other hand, would be more concerning. Such a condition can be ensured to be satisfied if $h''_{N}\geq 0$ throughout the physical domain, and this can be achieved for a sufficiently large value of the parameter $C$. Indeed, note that
	\begin{equation}
		h''_N(\psi) = 2C + \dv[2]{\psi}\left(\frac{\psi P_N(\psi)}{1-\alpha\psi}\right)\,,
	\end{equation}
	where the last term is a continuous function on the interval $[0,\psi_0)$, with a divergence at its endpoint. Assuming $c_j\in(0,\psi_0)$ $\forall j=1,\hdots,N$, both $P_N(0)$ and $P_N(\psi_0)$ will be positive, and therefore we can ensure the existence of a finite lower bound,
	\begin{equation}
		L\equiv \underset{0\leq \psi<\psi_0}{\rm min}\dv[2]{\psi}\left(\frac{\psi P_N(\psi)}{1-\psi/\psi_0}\right)\,.
	\end{equation}
	Choosing $C>\max(0,-L/2)$ ensures $h''_N(\psi)>0$, and therefore such a theory will satisfy the effective NEC. Consequently,
	\begin{equation}
		h'_N(\psi) = 1 + \int_0^\psi h''_N(u)\dd u > 1\,,
	\end{equation}
	and the derivative $h_N'>0$ will be of definite sign, which is also a necessary condition for regularity. 
	
\begin{figure}[t]
		\centering
		\includegraphics[width=\linewidth]{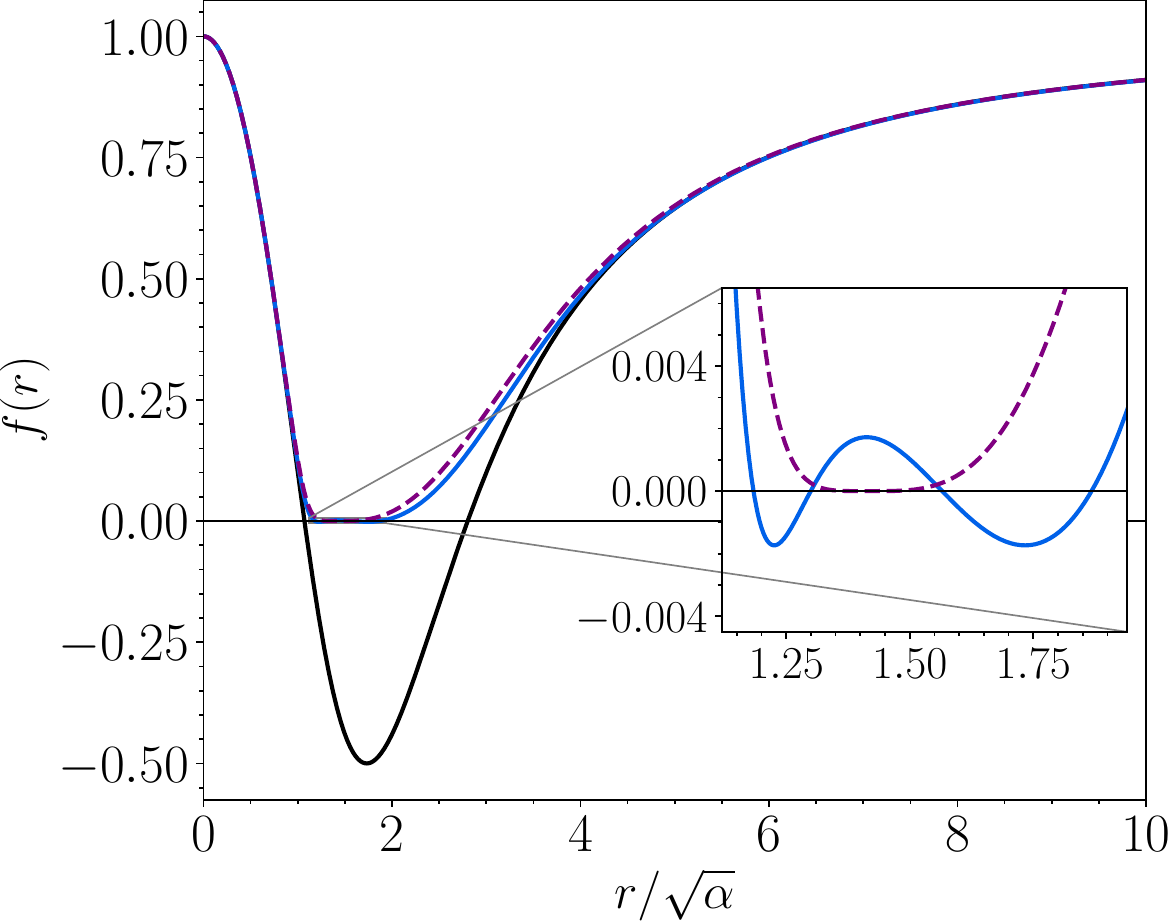}
		\caption{In black, metric function $f(r)$ for a Hayward spacetime~\eqref{eq:f_Hayward}, for $D=5$ spacetime dimensions and mass equal to $\mathsf M=4.5\alpha$. In blue, metric function for the solution of the deformed theory~\eqref{eq:hN}, with $N=2$ and parameters $c_1=1/(3\alpha)$, $c_2=2/(3\alpha)$ and mass $\mathsf M=4.515625\alpha$, displaying four different horizons---three inner horizons and one event horizon---. In purple, dashed line, same theory with parameters $c_1=c_2=1/(2\alpha)$ and mass $\mathsf M=4.5\alpha$, displaying a quadruple-degenerate horizon located precisely at $r_*=\sqrt{2\alpha}$. Both the blue and the purple curves correspond to $C=9\alpha$.}
		\label{fig:multiple_horizon}
	\end{figure}

	Such a multiple-horizon behavior can be seen in Fig.~\ref{fig:multiple_horizon}, compared to the case of Hayward, and for the particular case of $D=5$. The blue curve represents a particular case of the theory \eqref{eq:hN}, for parameters $N=2$, $c_1=1/(3\alpha)$ and $c_2=2/(3\alpha)$, depicting a four-horizon regular black hole, consisting of three inner horizons and one event horizon. Since the number of horizons of such a black hole is not bounded from above, we can have extended notions of extremality beyond the conventional extremal situation, and have four or more horizons merging in an \textit{ultra-extremal} scenario. For instance, this four-horizon black hole can be continuously deformed into an ultra-extremal black hole, where all four horizons merge at a single radius, as can be seen in Fig.~\ref{fig:multiple_horizon} (purple dashed line).
	
	\subsubsection{Near-horizon geometry of ``ultra-extremal'' black holes}
	
	We shall use the term \textit{ultra-extremal} to refer to black holes possessing a triple-degenerate horizon or a horizon of higher degeneracy.\footnote{
		Higher-order degenerate horizons in QT theories have been considered previously~\cite{DiFilippo:2024mwm,Hod:2024shg}, where the term \textit{super-extremal} was used. Since this term commonly refers instead to configurations beyond the extremality bound, such as $|Q|>M$ in Reissner--Nordstr\"om spacetimes, we adopt the term \textit{ultra-extremal}. The term \textit{ultracold} is also used for the triple-degenerate limit of Reissner-Nordstr\"om-de Sitter black holes~\cite{Romans1992,MannRoss1995}, but here we will not distinguish between those and horizons of higher degeneracy.
	}
	
	For SSS spacetimes satisfying $g_{tt}g_{rr}=-1$, as given in~\eqref{eq:ds2_SSS_N1}, conventional extremal horizons have a near-horizon geometry of the form (A)dS$_2\cross S_0^{D-2}$, where $S_0^{D-2}$ denotes the geometry of a $(D-2)$-sphere of radius $r_0$, \textit{i.e.,} the horizon radius. The $(t,r)$ sector is AdS$_2$ for $f''(r_0)>0$, or dS$_2$ for $f''(r_0)<0$. 
	
	For triple-degenerate black holes, where three horizons merge into a triple-degenerate horizon, the corresponding near-horizon limit instead yields ${\rm Mink}_2\cross S_0^{D-2}$, with ${\rm Mink}_2$ denoting two-dimensional Minkowski space~\cite{Castro:2022cuo}.
	
	Since the number of horizons is not bounded from above, it is natural to ask how this structure extends to horizons of arbitrary degeneracy. As we will see, horizons of degeneracy $n>3$ share an important property with the triple-degenerate case: the two-dimensional sector becomes locally flat as the horizon is approached. However, the way in which this flat limit is reached depends on the degeneracy order $n$.
	
	Let us consider an $n$-fold degenerate horizon located at $r=r_0$, so that the metric function can be expanded into
	\begin{equation}
		f(r)\sim c_n(r-r_0)^n
		+\hdots\,,\quad c_n\equiv \frac{f^{(n)}(r_0)}{n!}\neq0\,.
	\end{equation}
	To study the near-horizon geometry, it is convenient to introduce ingoing Eddington--Finkelstein coordinates through
	\begin{equation}
		\dd v = \dd t + \frac{\dd r}{f(r)}\,,
	\end{equation}
	in terms of which the metric takes the regular form
	\begin{equation}
		\dd s^2 = -f(r)\dd v^2 + 2\dd v \dd r + r^2\dd\Omega_{D-2}^2\,.
	\end{equation}
	We now perform the near-horizon scaling
	\begin{equation}
		r = r_0 + \epsilon \rho\,,\quad v = \frac{\tau}{\epsilon}\,,
	\end{equation}
	chosen such that the cross term $\dd v \dd r$ remains finite. The metric then behaves, to leading order in $\epsilon$, as
	\begin{equation}
		\dd s^2\big|_n \sim -c_n \epsilon^{n-2}\rho^n\dd \tau^2 + 2\dd\tau\dd\rho + (r_0+\epsilon\rho)^2\dd\Omega_{D-2}^2\,.
		\label{eq:ds2_nh}
	\end{equation}
	For $n=2$, both terms in the two-dimensional sector survive the $\epsilon\to0$ limit. Introducing a new coordinate $t$ through
	\begin{equation}
		\tau = t-\frac{1}{c_2\rho}
	\end{equation}
	the two-dimensional sector can be diagonalized,
	\begin{equation}
		\dd s^2_{(2)}\big|_{n=2} = -c_2\rho^2\dd t^2 + \frac{\dd \rho^2}{c_2\rho^2}\,.
	\end{equation}
	For $c_2\equiv 1/L_2^2>0$, this two-dimensional metric can be written in Poincaré coordinates of AdS$_2$ as follows,
	\begin{equation}
		\dd s_{(2)}^2
		=
		\frac{L_2^2}{z^2}
		\left(-\dd t^2+\dd z^2\right),\quad z\equiv \frac{1}{c_2\rho}\,.
	\end{equation}
	Therefore, the near-horizon geometry of doubly degenerate extremal black holes is locally AdS$_2\cross S_0^{D-2}$ in the case $f''(r_0)>0$. On the other hand, for $c_2<0$, the near-horizon two-dimensional geometry can be written in planar/conformal coordinates of dS$_2$ as follows,
	\begin{equation}
		\dd s^2_{(2)}\big|_{n=2} = \frac{L_2^2}{\eta^2}\left(\dd t^2 - \dd\eta^2\right),\quad \eta\equiv \frac{1}{|c_2|\rho}\,,
	\end{equation}
	where $c_2=-1/L_2^2$, with $L_2$ the dS$_2$ radius; note that $\eta$ is now the timelike coordinate, as $f<0$ on both sides of $r_0$. Therefore, for $n=2$ and $f''(r_0)<0$, the near-horizon geometry becomes locally dS$_2\cross S_0^{D-2}$, as expected~\cite{Castro:2022cuo}. Notice, however, that for $n>2$ the first piece of metric~\eqref{eq:ds2_nh} vanishes in the $\epsilon\to 0$ limit, yielding
	\begin{equation}
		\dd s^2\big|_{n>2} \to 2\dd\tau\dd\rho + r_0^2\dd\Omega_{D-2}^2\,,
	\end{equation}
	which can be written as Mink$_2\cross S_0^{D-2}$ by defining the pair of coordinates
	\begin{equation}
		(T,X)\equiv \left(\frac{\tau-\rho}{\sqrt{2}},\frac{\tau+\rho}{\sqrt{2}}\right)\,,
	\end{equation}
	yielding
	\begin{equation}
		\dd s^2\big|_{n>2} \to -\dd T^2 + \dd X^2 + r_0^2\dd\Omega^2_{D-2}\,,
	\end{equation}
	in the strict near-horizon limit $\epsilon\to 0$. Therefore, the near-horizon geometry of triple-degenerate or higher-degenerate (ultra-extremal) black holes is exactly Mink$_2\cross S_0^{D-2}$, in the $\epsilon\to 0$ limit, and hence the two-dimensional $(T,X)$ slices become flat.\footnote{
		An equivalent way of seeing that the near-horizon geometry becomes flat is checking the two-dimensional Ricci scalar, which for the two-dimensional $(t,r)$ sector of~\eqref{eq:ds2_SSS_N1} reads ${}^{(2)}R= -f''(r)\propto (r-r_0)^{n-2}$, and vanishes as we approach a degenerate horizon with $n>2$. Since in two dimensions the Riemann tensor is completely determined by the Ricci scalar, all polynomial curvature invariants constructed from the two-dimensional Riemann tensor vanish in this limit. The curvature of the $(t,r)$ sector therefore vanishes at any horizon of degeneracy $n>2$. 
	} Keeping the leading correction for small but finite \(\epsilon\), the two-dimensional metric reads
	\begin{equation}
		\begin{aligned}
			\dd s^2_{(2)}\big|_{n>2} &= -\left(1+\epsilon^{n-2}\frac{\tilde c_n(T,X)}{2}\right)\dd T^2 \\
			&+ \left(1-\frac{\epsilon^{n-2}}{2}\tilde c_n(T,X)\right)\dd X^2 \\
			&- \epsilon^{n-2}\tilde c_n(T,X)\dd T\dd X + \order{\epsilon^{n-1}}\,,
		\end{aligned}
	\end{equation}
	where we have defined,
	\begin{equation}
		\tilde c_n(T,X) \equiv \frac{c_n}{2^{n/2}}(X-T)^n\,.
	\end{equation}
	This metric is a small deformation of two-dimensional Minkowski spacetime, where $\epsilon^{n-2}|\tilde c_n|\ll 1$.
	
	Interestingly, although the limiting local geometry is the same for all $n>2$, the approach to this flat limit retains information about the degeneracy order, since different values of $n$ correspond to different scaling laws near the horizon. Ultra-extremal horizons are therefore locally indistinguishable at the level of the limiting curvature, but their near-horizon approach still carries information about the number of horizons that have merged.

	\subsection{More general theories}
	
	The pleasant analytical properties of QT gravities on spherical backgrounds \cite{Bueno:2024eig,Bueno:2024zsx} are due to the theories becoming specific two-dimensional Horndeski  theories \cite{Horndeski:1974wa} when reduced on their spherical sector. A general two-dimensional Horndeski theory, featuring a two-dimensional Lorentzian metric $\gamma_{\mu \nu}$ and a scalar field $\varphi$, takes the form:
	\begin{equation}
		\begin{split}
			\mathcal L_{\rm 2d} &= G_2(\varphi,X) - \square_\gamma \varphi G_3(\varphi,X) + G_4(\varphi,X)R^{\rm 2d} \\
			&- 2G_{4,X}(\varphi,X)\left[(\square_\gamma \varphi)^2-\nabla_\mu\nabla_\nu\varphi\nabla^\mu\nabla^\nu\varphi\right]\,,
		\end{split}
		\label{eq:horngen}
	\end{equation}
	where $\nabla$ stands for the two-dimensional covariant derivative, $\Box_\gamma =\gamma^{\mu \nu }\nabla_\mu \nabla_\nu$, $R^{\rm 2d}$ is the two-dimensional Ricci scalar, $X=(\partial \varphi)^2$ and the functions $\{G_i(\varphi,X)\}_{i=2}^4$ are arbitrary. Also, we use the notation $G_{i,\varphi} \equiv \partial_\varphi G_i$ and $G_{i,X} \equiv \partial_X G_i$. It should be noted that $G_4$ may be removed through appropriate partial integrations, so that $\mathcal L_{\rm 2d} $ is classically equivalent to the following theory $\bar{\mathcal{L}}_{\rm 2d}$:
	\begin{align}
		\label{eq:qthornintgen}
		\bar{\mathcal{L}}_{\rm 2d}&= \bar{G}_2(\varphi,X) - \square_\gamma \varphi \bar{G}_3(\varphi,X) \,, \\
		\bar{G}_2&=G_2+X \mathcal{I}_{,\varphi}\,, \quad  \bar{G}_3=G_3-2 G_{4,\varphi}- \mathcal{I}\,, \\& 
		\mathcal{I}=\int^X \frac{\mathrm{d} Y}{Y}  G_{4,\varphi}(\varphi,Y)\,.
		\label{eq:hornsim2gen}
	\end{align}
	Observe that $\mathcal{I}$ is determined only up to a function of $\varphi$, which would just contribute as a total derivative in~\eqref{eq:qthornintgen} and we can simply ignore it.

	Now, consider the following spherically symmetric $D$-dimensional metric:
	\begin{equation}
		\dd s^2 = \gamma_{\mu\nu}\dd x^\mu \dd x^\nu + \varphi(x)^2 \dd \Omega^2_{D-2}\,,
		\label{eq:spherred}
	\end{equation}
	where $\gamma_{\mu\nu}$ is a two-dimensional Lorentzian metric depending on the $\{x^\mu\}_{\mu=0}^1$ coordinates and $\varphi(x)$ is a scalar field on this lower-dimensional space. It can be shown \cite{Bueno:2024eig,Bueno:2024zsx} that the two-dimensional reduction of a general QT theory \eqref{eq:S_polyQTG} on \eqref{eq:spherred} gives rise to a two-dimensional Horndeski theory \eqref{eq:horngen} with functions $G_i=G_i^{\rm QT}$ given by 
	\begin{align}
		G_2^{\rm QT}(\varphi,X) &= \varphi^{D-2}[(D-1)h(\psi) - 2\psi h'(\psi)]\,, \\
		G_3^{\rm QT}(\varphi,X) &= 2\varphi^{D-3}h'(\psi)\,, \\
		G_4^{\rm QT}(\varphi,X) &= -\frac{\varphi^{D-2}}{2}\psi^{(D-2)/2}\int \dd\psi\frac{h'(\psi)}{\psi^{D/2}}\,,
	\end{align}
	where $\psi \equiv \dfrac{1-X}{\varphi^2 }$ and $h(\psi)$ is the so-called \textit{characteristic function} defined as in \eqref{eq:def_h}. The following relations hold:
	\begin{equation}
		G_{2,X}^{\rm QT}=-\frac{1}{2} G_{3,\varphi}^{\rm QT}\,, \quad G_3^{\rm QT}=2 G_{4,\varphi}^{\rm QT}\,.
	\end{equation}
	By the above, the two-dimensional reduction of QT gravities is classically equivalent to a theory $\bar{\mathcal{L}}_{\rm 2d}^{\rm QT}$, which takes the form of \eqref{eq:qthornintgen} with $\bar{G}_i=\bar{G}_i^{\rm QT}$ given by
	\begin{equation}
		\bar{G}_2^{\rm QT}=G_2^{\rm QT}+X \mathcal{I}_{,\varphi}^{\rm QT}\,, \quad  \bar{G}_3^{\rm QT}=G_3^{\rm QT}-2 G_{4,\varphi}^{\rm QT}- \mathcal{I}^{\rm QT}\,,
		\label{eq:gqtdef}
	\end{equation}
	with $\mathcal{I}^{\rm QT}$ given by substituting in \eqref{eq:hornsim2gen} $G_4 \rightarrow G_4^{\rm QT}$. Note that
	\begin{equation}
		\bar{G}_{2,X}^{\rm QT}=-\bar{G}_{3,\varphi}^{\rm QT}\,.
	\end{equation}
	
	Take the following Weyl transformation of the metric:
	\begin{equation}
		\label{eq:redef}
		\gamma_{\mu \nu}=\Omega^2(\varphi, Z) \lambda_{\mu \nu}\,, \quad Z=\lambda^{\mu \nu} \partial_\mu \varphi \partial_\nu \varphi\,,
	\end{equation}
	where the Weyl rescaling $\Omega>0$ is allowed to depend on $\varphi$ and $Z$. We further assume $\Omega$ is bounded on the relevant domain $\mathcal{D}_Z \subset \mathbb{R}^2$ and the transformation is invertible, which requires that 
	\begin{equation}
		\left ( \Omega-2 Z \partial_Z \Omega \right ) \vert_{(\varphi_0,Z_0)} \neq 0\,, \quad \forall \, (\varphi_0,Z_0) \in \mathcal{D}_Z\,.
		\label{eq:nonzerjaco}
	\end{equation}
	For global statements about event horizons, we will also assume that the transformation identifies the corresponding conformal boundaries and spacetime extensions. In the asymptotically flat case, for example, choosing some coordinates $(t,r)$ and fixing the gauge $\varphi=r$ so that $\lim_{r \rightarrow \infty} \gamma_{\mu \nu}=\lim_{r \rightarrow \infty}\lambda_{\mu \nu}=\eta_{\mu \nu}$, the Weyl factor satisfies $\lim_{r \rightarrow +\infty} \Omega(r,1)=1$. 
	
	Under these conditions, the Weyl rescaling \eqref{eq:redef} preserves the causal structure of the two-dimensional Lorentzian manifold, so that the horizon structure of the metrics $\gamma_{\mu \nu}$ and $\lambda_{\mu \nu}$ will be the same. As a result, if one considers the transformed two-dimensional theory $\bar{\mathcal{L}}'_{\rm 2d}$ in terms of $(\lambda_{\mu \nu}, \varphi)$,
	\begin{align}
		\sqrt{\vert \gamma \vert}\bar{\mathcal{L}}_{\rm 2d}^{\rm QT}(\gamma_{\mu \nu},\varphi)&= \sqrt{\vert \lambda \vert} \bar{\mathcal{L}}'_{\rm 2d}(\lambda_{\mu \nu},\varphi)\,, \\ 
		\bar{\mathcal{L}}'_{\rm 2d}=\Omega^2  \bar{G}_2^{\rm QT} (\varphi& ,\Omega^{-2} Z)- \Box_\lambda  \varphi \bar{G}_3^{\rm QT}(\varphi,\Omega^{-2} Z)\,,
	\end{align}
	it turns out that a solution $(\gamma_{\mu \nu},\varphi)$ of $\bar{\mathcal{L}}_{\rm 2d}^{\rm QT}$ will be mapped into a solution $(\lambda_{\mu \nu},\varphi)$ of $\bar{\mathcal{L}}'_{\rm 2d}$ with identical causal structure. Therefore, the results we have obtained in this manuscript regarding the horizon structure of solutions of QT gravities\footnote{Observe that we reserve the name of QT gravities to those higher-curvature theories constructed from polynomial contractions of curvature tensors without covariant derivatives (possibly including an infinite tower of them) with second-order equations of motion on spherical backgrounds.} will also apply to more general higher-derivative (non-polynomial) theories whose spherical sector produces a Horndeski theory related by a Weyl rescaling \eqref{eq:redef} to another Horndeski theory classically equivalent to that defined by \eqref{eq:gqtdef}. In fact, on the domain $\mathcal{D}_Z$ for which \eqref{eq:nonzerjaco} holds, the transformation \eqref{eq:redef} gives a one-to-one correspondence between solutions of $\bar{\mathcal{L}}_{\rm 2d}^{\rm QT}$ and $\bar{\mathcal{L}}_{\rm 2d}'$.
	
	Moreover, note that there always exists a point $(\varphi_0,X_0)$ on $\mathcal{D}_X=\{(\varphi,X) \, \vert\,  (\varphi,\Omega^{2} X)\in \mathcal{D}_Z \}$ such that
	\begin{equation}
		\bar{G}_2^{\rm QT}-X \bar{G}_{2,X}^{\rm QT} \vert_{(\varphi_0,X_0)}=G_2^{\rm QT} \vert_{(\varphi_0,X_0)} \neq 0\,,
	\end{equation}
	as we assume that $G_2^{\rm QT}$ does not vanish identically on $\mathcal{D}_Z$. Therefore, for any function $\mathcal{G}_2(\varphi,Z)$ for which there exists $\Omega_0>0$ such that $\mathcal{G}_2(\varphi_0,\Omega_0^2 X_0)=\Omega_0^2 \bar{G}_2^{\rm QT}(\varphi_0,X_0)$ and $\mathcal{G}_2(\varphi_0,\Omega_0^2 X_0) \neq \Omega_0^2 X_0 \mathcal{G}_{2,Z}(\varphi_0,\Omega_0^2 X_0)  $ (this comes from \eqref{eq:nonzerjaco}), the following equation
	\begin{equation}
		\label{eq:defomega}\mathcal{G}_2(\varphi,Z)=\Omega^2 \bar{G}_2^{\rm QT}(\varphi,\Omega^{-2} Z)
	\end{equation}
	defines implicitly $\Omega(\varphi,Z)$ locally around a point\footnote{Note that $G_2^{\rm QT}$ would vanish at a degenerate horizon for extremal black holes. However, even in this case, the causal structure remains unmodified with respect to the original solution of the starting QT theory. For extremal black holes, the equation \eqref{eq:defomega} defines implicitly $\Omega$ locally around a point away from the degenerate horizon.}   $(\varphi_0,\Omega_0^{2} X_0)$ in which $\Omega(\varphi_0,\Omega_0^{2} X_0)=\Omega_0$.  Once the Weyl factor $\Omega$ is fixed by \eqref{eq:defomega}, we have that
	\begin{equation}
		\bar{\mathcal{L}}'_{\rm 2d}=\mathcal{G}_2(\varphi,Z)- \Box_\lambda \varphi \mathcal{G}_3(\varphi,Z)\,, 
	\end{equation}
	where $\mathcal{G}_3(\varphi,Z)=\bar{G}_3^{\rm QT}(\varphi,\Omega^{-2} Z)$. As a result, we have been able to show that solutions of QT gravities capture the horizon structure of a far more general class of higher-derivative theories, which reduce to different two-dimensional Horndeski theories on top of spherically symmetric backgrounds.

	\section{The space of regular quasi-topological theories}\label{sec:space}       
	The space of theories that yield regular black hole solutions (or any regular spacetime, in general) corresponds to the subset of QT theories whose solutions satisfy the regularity conditions, which arise from imposing bounded curvature invariants. For spherically symmetric spacetimes of the form~\eqref{eq:ds2_SSS_N1}, such regularity conditions, in terms of the function $\psi(r)$ were given in~\eqref{eq:conds_psi}.
	
	As we have seen, if this spacetime is a solution of a QT theory, we can write $\psi(r) = h^{-1}[S(r)]$, where $h^{-1}$ is the inverse of the characteristic function, and $S(r)$ is defined as the right-hand side of the algebraic equation~\eqref{eq:qtgSS_eqs_2}. Let us write the inverse function as $h^{-1}\equiv H$. Note that the first regularity condition implies $h(\psi)$ diverges at the constant finite value $\psi=\psi_0$, which means that the image of the inverse function $h^{-1}$ is bounded. Notice that the location of the endpoint of the domain of $h$ can be fixed by a simple rescaling. Indeed, let
	\begin{equation}
		p\equiv \psi/\psi_0\,,\quad \tilde h(p)\equiv \frac{1}{\psi_0}h(\psi)\,,\quad p\in[0,1)\,,
		\label{eq:rescalings}
	\end{equation}
	which places the endpoint at $p=1$. This transformation preserves the normalization of the Einstein term, since
	\begin{equation}
		\tilde h'(0)=h'(0)=1\,,
	\end{equation}
	and the constant term transforms as
	\begin{equation}
		\tilde h(0)=\frac{h(0)}{\psi_0}
		=-\frac{2\Lambda}{(D-1)(D-2)\psi_0}\,.
	\end{equation}
	Thus, for asymptotically flat solutions, $\tilde h(0)=0$, whereas for asymptotically (A)dS solutions the bare cosmological constant is expressed in units of the core curvature scale. Moreover,
	\begin{equation}
		\tilde h'(p) = h'(\psi_0 p)\,,\quad \tilde h''(p) = \psi_0 h''(\psi_0 p)\,,
	\end{equation}
	so strict monotonicity is unchanged. Since $h(\psi)$ diverges as $\psi\to\psi_0^-$, one also has the same behavior $\tilde h(p)\to +\infty$ when $p\to 1^-$. Therefore, the endpoint regularity conditions are preserved up to an irrelevant constant factor. Indeed,
	\begin{equation}
		\frac{\tilde h}{\tilde h'} = \frac{1}{\psi_0}\frac{h}{h'}\,,\quad \frac{\tilde h^2\tilde h''}{(\tilde h')^3} = \frac{1}{\psi_0}\frac{h^2h''}{(h')^3}\,,
	\end{equation}
	where the expressions at the right-hand side are calculated at $\psi = \psi_0 p$. We may therefore place the endpoint at $\psi=1$ without loss of generality, and drop the tildes from now on for the sake of the notation.
	
	Denoting $\psi = H(S)$, such conditions can be rewritten into the following form,
	\begin{equation}
		\lim_{S\to\infty} SH'(S) = \lim_{S\to\infty} S^2H''(S) = 0
		\label{eq:reg_conds_1}
	\end{equation}
	for analytic functions
	$H(S)$ with bounded image, together with
	\begin{equation}
		\lim_{S\to\infty} H(S) = 1\,.
		\label{eq:reg_conds_2}
	\end{equation}
	Therefore, we can write the regularity conditions~\eqref{eq:reg_conds_1} and~\eqref{eq:reg_conds_2} in the following compact form,
	\begin{equation}
		\lim_{S\to\infty}S^n H^{(n)}(S) = \delta_{n0}\,,\quad \text{for}\quad n=0,1,2\,,
		\label{eq:reg_conds_3}
	\end{equation}
	where $\delta_{n0}$ is the Kronecker delta. Therefore, any regular spacetime that can be obtained as a QT vacuum solution (not belonging to the degenerate branch) will be solely described by a single function $f(r)$, satisfying $g_{tt}g_{rr}=-1$, and such that
	\begin{equation}
		f(r) = 1-r^2H(S(r))\,.
	\end{equation}
	For the sake of simplicity, let us consider asymptotically flat spacetimes and, correspondingly, characteristic functions such that $h(0)=H(0)=0$. We define the \textit{unrestricted inverse function space} $\mathcal{H}_0$ as follows:
	\begin{align}
		\notag
		\mathcal H_0 \equiv \{&H(S):[0,\infty)\to[0,1)\mid H(0)=0\,, H'(0)=1\, , \\ &H'(S) >0\;\text{and}\;
		\text{\eqref{eq:reg_conds_3} are satisfied}\}\,,
		\label{eq:inverse_func_space}
	\end{align}
	where the first condition encodes asymptotic flatness, the second one preserves the normalization of the Einstein gravity term, and the third one corresponds to the monotonicity requirement. Observe that $\mathcal{H}_0$ may include functions which do not correspond to the inverse of \emph{physical} characteristic functions $h(\psi)$, as in even dimensions $D=2n$ one must require that $h^{(n)}(0)=0$---this comes from the fact that $n$-th order QT gravities for $D=2n$ do not contribute to the equations of motion on spherical symmetry. As a result, depending on the spacetime dimension $D$, we define the \emph{regular characteristic function space} $\mathcal{C}_0^{(D)}$ as:
	\begin{align}
		\label{eq:charac_func_space1}
		\mathcal{C}_0^{(2n+1)} & \equiv
		\{h:[0,1)\to [0,\infty)\mid h^{-1}\in\mathcal{H}_0\}\,,
		\\ \mathcal{C}_0^{(2n)} & \equiv
		\{h:[0,1)\to [0,\infty)\mid h^{-1}\in\mathcal{H}_0 \, ,\, h^{(n)}(0)=0 \}\,.
		\label{eq:charac_func_space2}
	\end{align}

Although such functional spaces are not vector spaces, these can be regarded as subsets of the vector space of real analytic functions, $C^\omega([0,1))$. 
For the following arguments, we do not fix a particular topology on these spaces, but assume that the operations entering the deformations below are continuous, including linear combinations, functional inversion and the derivative evaluations required in even dimensions.

\subsection{Contractibility of the regular characteristic function space}\label{sec:space_contractibility}

The way the characteristic function space $\mathcal{C}_0^{(D)}$ is defined in~\eqref{eq:charac_func_space1} and \eqref{eq:charac_func_space2}  may be used to prove that it is contractible and, in particular, path-connected and simply connected (\textit{i.e.} every loop in $\mathcal C_0^{(D)}$ can be continuously shrunk to a point). In order to prove this result, we will show that the identity map on $\mathcal{C}_0^{(D)}$ is homotopic to a constant map on  $\mathcal{C}_0^{(D)}$. To this aim, let $h_0 \in \mathcal{C}_0^{(D)}$ be arbitrary and let $h_1 \in \mathcal{C}_0^{(D)}$ be a fixed characteristic function. Define  $H_0=h^{-1}_0$ and $H_1=h_1^{-1}$. Consider the following one-parameter family of functions $H_\lambda \in \mathcal{H}_0$ interpolating between $H_0$ and $H_1$\,,
\begin{equation}
	H_\lambda \equiv \lambda H_0+(1-\lambda) H_1\,, \quad \lambda \in [0,1]\,.
\end{equation}	
Observe that $H_\lambda(0)=0$ and $H_\lambda'(0)=1$. Let $\tilde{h}_\lambda=H^{-1}_\lambda$ for $\lambda \in [0,1]$. By construction, $\tilde{h}_\lambda(0)=0$, $\tilde{h}_\lambda'(0)=1$ and
\begin{equation}
	\tilde{h}_\lambda\to+\infty,\qquad
	\frac{\tilde{h}_\lambda}{\tilde{h}_\lambda'}\to0,\qquad
	\frac{\tilde{h}_\lambda^2\tilde{h}_\lambda''}{(\tilde{h}_\lambda')^3}\to0
	\qquad (x\to1^-).
	\label{eq:reg_conds_hlambda}
\end{equation}
for any $\lambda \in [0,1]$. For odd dimensions, $\tilde{h}_\lambda \in \mathcal{C}_0^{(2n+1)}$ and we have constructed a continuous map from the identity map to a constant one, thus proving that $\mathcal{C}_0^{(2n+1)}$ is contractible.  However, for even $D=2n$ with $n \geq 2$, $\tilde{h}_\lambda$ does not need to belong to $\mathcal{C}_0^{(2n)}$ if $\lambda \in (0,1)$. We may fix this by subtracting an appropriate function from $\tilde{h}_\lambda$ in even dimensions, so as to preserve the defining conditions of $\mathcal C_0^{(2n)}$ given in~\eqref{eq:charac_func_space2}.
Indeed, define the auxiliary function,
\begin{equation}
	A_\lambda(y)\equiv \beta_\lambda y^n e^{-\gamma_\lambda y}\,,
\end{equation}
where the parameters $\beta_\lambda$ and $\gamma_\lambda$ are defined as follows,
\begin{equation}
	\beta_\lambda = \frac{\tilde h_\lambda^{(n)}(0)}{n!}\,,\quad \gamma_\lambda^{n-1} = 2n!|\beta_\lambda|\,.
\end{equation}
Now, define 
\begin{equation}
	h_\lambda(\psi) = \tilde h_\lambda(\psi) - A_\lambda\left[\tilde h_\lambda(\psi)\right]\,.
	\label{eq:finalinterpola}
\end{equation}
Note that, when $\beta_\lambda = 0$, the correction vanishes. Since $\tilde h_\lambda(\psi) = \psi + \order{\psi^2}$, its expansion around the origin gives
\begin{equation}
	A_\lambda\left[\tilde h_\lambda(\psi)\right] = \frac{\tilde h_\lambda^{(n)}(0)}{n!}\psi^n + \order{\psi^{n+1}}\,,
\end{equation}
and therefore we obtain $h_\lambda^{(n)}(0)=0$ for any $\lambda\in[0,1]$, while $h_\lambda(0)=0$ and $h'_\lambda(0)=1$ are still preserved. The choice of $\gamma_\lambda$ ensures this correction also preserves the monotonicity. Indeed, for $\beta_\lambda\neq 0$, setting $x=\gamma_\lambda y$ yields
\begin{equation}
	\left|\dv{A_\lambda}{y}\right| = \frac{|\beta_\lambda|}{\gamma_\lambda^{n-1}}x^{n-1}e^{-x}|n-x|\leq \frac{n!|\beta_\lambda|}{\gamma_\lambda^{n-1}} = \frac{1}{2}\,.
\end{equation}
The inequality can be straightforwardly checked using the series expansion of the exponential function. Therefore,
\begin{equation}
	h'_\lambda(\psi) = \tilde h'_\lambda(\psi)\left(1-A'_\lambda\left[\tilde h_\lambda(\psi)\right]\right)\geq \frac{1}{2}\tilde h'_\lambda > 0\,.
\end{equation}
For nonzero $\beta_\lambda$, the auxiliary function and its derivatives decay exponentially in the endpoint limit $\psi\to 1^-$, and therefore $h_\lambda\sim \tilde h_\lambda$, directly satisfying the regularity conditions in~\eqref{eq:reg_conds_hlambda}. These conditions are also unchanged when $\beta_\lambda=0$, since the correction then vanishes. Hence, $h_\lambda\in\mathcal C_0^{(2n)}$ for every $\lambda\in[0,1]$, as we wanted to show. Finally, $\beta_0=\beta_1=0$, so the endpoints of the interpolation remain unchanged. Both parameters depend continuously on information of the Taylor series of $\tilde h_\lambda$ at the origin, and the correction and all its derivatives vary continuously along admissible families of theories.

By the assumptions above, we have been able to construct a continuous map from the identity map to a constant map.  Therefore, we prove that $\mathcal{C}_0^{(2n)}$ is also contractible\footnote{
	As  contractible spaces have all homotopy groups trivial, in particular $\pi_1\left (\mathcal{C}_0^{(D)} \right)=0$ and then $\mathcal{C}_0^{(D)}$ is simply connected too (in particular, path-connected).
	\label{ft:topology2}
} for even spacetime dimensions. 

The preceding arguments can also be formulated directly in terms of continuously parametrized families.
Indeed,  contractibility only involves families of characteristic functions continuously parametrized by compact Hausdorff spaces, namely the interval $[0,1]$, the circle $S^1$ and their products. This is precisely the structure of a \textit{quasi-topological space} in the sense of Spanier~\cite{Spanier:1963}: a set $X$ together with a prescription of which maps $K\to X$, with $K$ compact Hausdorff, are regarded as continuous, subject to natural compatibility conditions. A natural quasi-topology on $\mathcal C_0^{(D)}$ is obtained by declaring a family $\{h_k\}_{k\in K}\subset\mathcal C_0^{(D)}$ admissible whenever $h_k(\psi)$ and all its derivatives are jointly continuous in $(k,\psi)\in K\times[0,1)$. Since the interpolation \eqref{eq:finalinterpola} maps admissible families into admissible families, we find that  $\mathcal{C}_0^{(D)}$ is a  contractible quasi-topological space. In other words, the space of quasi-topological theories admitting regular black holes is, quite literally, a contractible quasi-topological space.

In conclusion, the results of this section imply that the space of QT theories admitting RBH solutions is not only connected, but also path-connected, simply connected and contractible, meaning that any theory with RBHs may be continuously deformed into any other one without ever encountering a theory admitting singular black holes. Hence the space $\mathcal{C}_0^{(D)}$ has a single connected component and features no \emph{islands} of QT theories admitting RBHs. 

\subsection{Stability under small coupling deformations}\label{sec:space_stability}

So far, we have shown that the regular characteristic function space
$\mathcal{C}_0^{(D)}$, defined in~\eqref{eq:charac_func_space1} and \eqref{eq:charac_func_space2}, is contractible and simply connected. That is, we showed that any two elements of this space can be continuously deformed into one another, following a path that is entirely contained within the space itself. This means we can, in principle, take two arbitrary theories that yield spherically symmetric regular black holes in the vacuum, and construct a one-parameter family of theories that interpolates between them while preserving the regularity of the solution. 

This result, however, does not establish whether regularity is preserved under small deformations of the theory. That is, given a theory admitting a regular black hole solution, do sufficiently small changes in its couplings preserve the regularity of its solution? To address this, consider a set of higher-order couplings $\{\alpha_n\}$ defining a theory whose characteristic function $h(\psi)$ belongs to $\mathcal{C}_0^{(D)}$. To this set of couplings we consider a perturbation,
\begin{equation}
	\{\alpha_n\} \longrightarrow
	\{\widetilde\alpha_n\}=\{\alpha_n+\delta\alpha_n\}\,,
\end{equation}
which  defines a new characteristic function
\begin{equation}
	\widetilde h(\psi)=h(\psi)+\delta h(\psi) \, .
\end{equation}
Our objective is to identify sufficient conditions on $\delta \alpha_n$ such that $\widetilde h(\psi)$ also belongs to $\mathcal{C}_0^{(D)}$.

To this end, we prove two complementary results. First, if all the couplings are nonnegative, we show the theory stays regular when each coupling is changed by a fraction of its own value. Second, relaxing any assumption on the signs of the couplings, we show that any regular theory can be deformed by increasing or decreasing each coupling, as long as certain weighted sums of these changes stay small. Thus, regularity is preserved under broad classes of deformations to the entire tower of higher curvature couplings.

\subsubsection{Subset of positive couplings $\alpha_n\geq 0$}
A particularly simple example arises whenever the couplings $\alpha_n$ defined in~\eqref{eq:an_alphan} are positive. As pointed out in~\cite{Bueno:2024dgm}, requiring all the higher-order couplings of the theory to be positive is part of the set of sufficient conditions for regularity. In particular, it ensures monotonicity of the characteristic function, $h'(\psi)>0$. Note, however, that this is not a necessary condition, as theories with some of the couplings negative can nevertheless satisfy the monotonicity condition. In this way, theories with positive higher-order couplings will correspond to a particular subset of the regular characteristic function space $\mathcal{C}_0^{(D)}$, which is also of particular interest, given that some known regular black holes, such as Hayward~\cite{Hayward:2005gi}, are vacuum solutions of theories that belong to this subset.\footnote{In this paragraph, when we mention positive couplings, rather than referring to the couplings that appear at the level of the action, $a_n$, we refer to the $\alpha_n$ defined in~\eqref{eq:an_alphan}.}

Within this particular subset of theories, one can show that \textit{small} perturbations in coupling space deform the characteristic function in such a way that the new theory also belongs to the regular characteristic function space, $\widetilde h = h + \delta h \in\mathcal C_0^{(D)}$. We regard \textit{small} perturbations as those satisfying
\begin{equation}
	|\delta\alpha_n|\leq \epsilon\alpha_n\,,
	\label{eq:suff_cond_coupling}
\end{equation}
for an arbitrary parameter $\epsilon\in(0,1)$. This condition, as we will see, is a sufficient condition for $\widetilde h \in\mathcal C_0^{(D)}$. Let us define the new couplings as $\widetilde\alpha_n = \alpha_n+\delta\alpha_n$. Then, from~\eqref{eq:suff_cond_coupling} we immediately get
\begin{equation}
	(1-\epsilon)\alpha_n \leq \widetilde\alpha_n \leq (1+\epsilon)\alpha_n\,.
\end{equation}
Therefore, for $x\in[0,1)$, we obtain
\begin{equation}
	(1-\epsilon)h^{(k)}(x)\leq \widetilde h^{(k)}(x)\leq (1+\epsilon)h^{(k)}(x)\,,
\end{equation}
for any derivative order $k$. Hence, if $h(x)\to+\infty$ in the $x\to 1^-$ limit, the new characteristic function will preserve this condition. The bound on the first derivative, $\widetilde h'(x)\geq (1-\epsilon)h'(x) > 0$ ensures monotonicity. The first derivative condition of~\eqref{eq:conds_h} reads
\begin{equation}
	\lim_{x\to1^-}\frac{\widetilde h(x)}{\widetilde h'(x)}\leq \frac{1+\epsilon}{1-\epsilon}\lim_{x\to 1^-}\frac{h}{h'} = 0\,,
\end{equation}
and similarly for the second derivative condition,
\begin{equation}
	\lim_{x\to 1^-}\frac{\widetilde h^2\widetilde h''}{(\widetilde h')^3} \leq \frac{(1+\epsilon)^3}{(1-\epsilon)^3}\lim_{x\to 1^-}\frac{h^2h''}{h'^3} = 0\,,
\end{equation}
which allows us to state this result in the following form: 

\textit{Theorem:} Given a regular theory with $\alpha_n\geq 0$, every deformation of the higher-order couplings ${\alpha_n}\to\{\alpha_n+\delta\alpha_n\}$, $n\geq2$, satisfying $|\delta\alpha_n|\leq\epsilon\alpha_n$ for an arbitrary parameter $\epsilon\in(0,1)$ preserves regularity.

It is important to emphasize, however, that condition~\eqref{eq:suff_cond_coupling} is sufficient but not necessary, and other notions of \textit{small} deformations may allow for other theories within the characteristic function space. For instance, for any regular theory (not necessarily with $\alpha_n\geq0$), a deformation of a finite number of couplings,\footnote{
	In even spacetime dimensions, admissible deformations must
	satisfy $\delta\alpha_{D/2}=0$, since the term proportional to $x^{D/2}$ is absent from the characteristic function of the
	quasi-topological theories considered here.
}
\begin{equation}
	\delta h(x) = \sum_{n=2}^N\delta\alpha_n x^n\,,
\end{equation}
preserves regularity, provided $\sum_n n|\delta\alpha_n|<{\rm inf}_{[0,1)}h'$. This infimum is positive, since $h'>0$ and $h'\to+\infty$ as $x\to 1^-$. The condition preserves monotonicity, and a polynomial is bounded, together with its derivatives, at $x=1$, so it does not affect the endpoint conditions~\eqref{eq:conds_h}.
		
		\subsubsection{Arbitrary couplings $\alpha_n$}
		As we have mentioned in the previous subsection, requiring $\alpha_n\geq0$ is a sufficient condition for $h'>0$, but not necessary, as theories with some of the couplings negative can very well satisfy the monotonicity condition. Without putting any restriction on the sign of the couplings, we can also prove a similar theorem. Consider a characteristic function $h(\psi)\in\mathcal C_0^{(D)}$ with higher-order couplings $\{\alpha_n\}$, for $n\geq2$. Suppose, for the sake of simplicity, that the solution of such a theory is asymptotically flat, $h(0)=0$. Since $h'>0$, we introduce $c\equiv \inf_{[0,1)}h'>0$. Now, consider a set of deformations given by
		\begin{equation}
			\delta h(\psi) = \widetilde h - h = \sum_{n\geq2}\delta\alpha_n \psi^n\,,\qquad |\delta\alpha_n| \leq \varepsilon_n\,,
			\label{eq:dh}
		\end{equation}
		and where the nonnegative parameters $\varepsilon_n$ are defined from the  following requirements
		\begin{equation}
			\sum_{n\geq2}n\varepsilon_n < c\,,\qquad \sum_{n\geq2}n(n-1)\varepsilon_n<\infty\,.
			\label{eq:requirements_epsn}
		\end{equation}
		Note that $\varepsilon_n$ has to decay faster than $1/n^3$ in the large $n$ limit for the second restriction to hold. A particular example can be $\varepsilon_n= c\,2^{-n}/n^2$. Such assumptions ensure that $\delta h$ and its first two derivatives are bounded on $[0,1)$. Moreover, 
		\begin{equation}
			\widetilde h'\geq h'-\sum_{n\geq2}n\varepsilon_n>0\,,
		\end{equation}
		so strict monotonicity is preserved. Since $h,h'\to+\infty$ and the perturbation remains bounded,
		\begin{equation}
			\frac{\widetilde h}{h}\to 1\,,\qquad\frac{\widetilde h'}{h'}\to 1\,.
		\end{equation}
		Thus $\widetilde h\to+\infty$ and $\widetilde h/\widetilde h'\to 0$. Finally,
		\begin{equation}
			\frac{\widetilde h^2\widetilde h''}{(\widetilde h')^3} = \frac{(\widetilde h/h)^2}{(\widetilde h'/h')^3}\left[\frac{h^2h''}{(h')^3}+\frac{h^2\delta h''}{(h')^3}\right]\to 0\,,
		\end{equation}
		because $\delta h''$ is bounded and
		\begin{equation}
			\frac{h^2}{(h')^3} = \left(\frac{h}{h'}\right)^3\frac{1}{h'}\to 0\,.
		\end{equation}
		Therefore, such results allow us to state it in the following form:
		
		\textit{Theorem:} For any given element $h\in\mathcal C_0^{(D)}$, corresponding to a theory that yields a unique vacuum and spherically symmetric solution, which is specified by its higher-order couplings $\{\alpha_n\}_{n\geq2}$, every deformation $\delta h = \widetilde h - h$ satisfying~\eqref{eq:dh} and~\eqref{eq:requirements_epsn} preserves regularity, and therefore $\widetilde h\in\mathcal C_0^{(D)}$.
		
		It is important to mention that for the even-dimensional case, we need to require $\delta\alpha_{D/2}=0$ in order for $\widetilde h\in\mathcal C_0^{(D)}$. This additional restriction does not modify the preceding arguments.
		
		\subsection{Theories admitting RBHs are dense}
		Another interesting characterization of the space of quasi-topological theories is that any theory defined by a set of couplings $\{\alpha_n\}_{n\geq2}^N$, up to any arbitrarily high, but finite, curvature order $N$, can be completed into a theory yielding a regular black hole as a solution, while preserving the same set of lower-order couplings. To understand why this is the case, one should note that the regularity conditions are actually requirements imposed on the divergences of $h$ and its derivatives. That is, regularity is determined by the tail of the characteristic function. Hence, one can argue that such theories admitting regular black holes are actually \textit{dense}, in the sense of coefficient convergence: any theory that yields a singular black hole can be approached by a sequence of theories admitting regular black holes that agree with it through an increasingly high curvature order. In particular, no finite set of couplings is sufficient to determine whether a theory admits regular black holes or not,
		since the remaining higher-order terms can be chosen to ensure regularity without modifying those couplings.
		
		To see this in more detail, let us consider the following example. Consider a $D$-dimensional theory of the quasi-topological class with finitely many corrections to the Einstein-Hilbert action, whose characteristic function reads
		\begin{equation}
			h_{N}(\psi) = \psi + \sum_{n=2}^N\alpha_n \psi^n\,,
			\label{eq:hN_example}
		\end{equation}
		and where we have assumed that the solution is asymptotically flat, $h(0)=0$, for simplicity, together with the restriction $\alpha_{D/2}=0$ for the even-dimensional case. Choose an odd integer $K>\max(N,D/2)$ and define
		\begin{equation}
			h(\psi) = h_{N}(\psi) + \frac{A}{\alpha}\frac{(\alpha\psi)^K}{1-\alpha\psi}\,,\quad A>0\,,
			\label{eq:h_completed_rbh}
		\end{equation}
		for some sufficiently large dimensionless coefficient $A>0$ chosen to ensure monotonicity $h'>0$, and where $\alpha>0$ is a parameter with units of length$^2$, which need not be related to the lower-order couplings $\alpha_n$ of~\eqref{eq:hN_example}. Note that the added term changes no coefficient through order $N$, it preserves the normalization $h'(0)=1$ and the admissibility in even-dimensional spacetimes, and it is monotonically increasing in the interval $\psi\in[0,1/\alpha)$. It is straightforward to check that such a theory satisfies the regularity conditions~\eqref{eq:conds_h}, and therefore it belongs to the previously defined regular characteristic function space $h\in\mathcal C_0^{(D)}$, in any spacetime dimension.
		
		Such a completion, however, is not unique, as we can choose the parameter $K$ freely, together with the shape of the tail. In general, such a completion takes the form $h(\psi)=h_N(\psi) + \psi^K \mathcal F(\psi)$ for some $K>\max(N,D/2)$ and arbitrary continuous and differentiable function $\mathcal F(\psi)$ on the interval $\psi\in[0,1/\alpha)$, that satisfies the regularity conditions~\eqref{eq:conds_h}, together with
		\begin{equation}
			K\frac{\mathcal F(\psi)}{\psi} + \mathcal F'(\psi)>-\psi^{-K}h'_N(\psi)\,,
		\end{equation}
		ensuring the corresponding $h(\psi)$ is monotonically increasing.
		
		\section{Conclusions}\label{sec:conclusions}
		In this work, we have investigated general properties of the space of quasi-topological gravity theories admitting pure-gravity RBHs, as well as the structure of their corresponding spaces of solutions.
		
		Our first main result is that every positive-mass regular vacuum solution within the class considered here necessarily possesses a de Sitter core. Minkowski and anti-de Sitter cores are excluded, except for the corresponding maximally symmetric vacuum spacetimes. Thus, the appearance of a positively curved core is not a special feature of particular regular black hole geometries, but a universal property of the vacuum solutions considered here.
		
		We also obtained general results concerning the horizon structure. Asymptotically flat and anti-de Sitter regular black holes necessarily admit at least one extremal configuration.
		Without imposing further restrictions on the characteristic function, we argued that there is no upper bound on the number of horizons, extremal or ultra-extremal configurations that a theory may admit, and provided a particular example in~\eqref{eq:hN}. By contrast, if all the couplings are nonnegative, the extremal
		configurations are unique, and there are at most two horizons for asymptotically flat or AdS solutions. We also showed that these results extend to a much broader class of theories whose SS sector is described by more general Horndeski two-dimensional actions.
		
		In addition, we identified a degenerate branch of theories for which $h(\psi)=h'(\psi)=0$. At such points, the usual uniqueness properties of the spherically symmetric vacuum solution are lost and a much larger set of geometries may arise, including redshifted Einstein static Universes and genuinely time-dependent vacua, although traversable wormholes remain excluded. These branches, however, require a special tuning of the couplings and are qualitatively different from the generic regular sector.
		
		A central result of this work is the characterization of the space of regular theories. We showed that the space of inverse characteristic functions satisfying the regularity conditions, as well as the corresponding space of characteristic functions, are contractible and, in particular, path-connected and simply connected. Therefore, regular theories do not appear as disconnected or isolated regions of theory space.
		
		We further showed that regularity is stable under small deformations of the higher-curvature couplings. In particular, within the sector with positive coefficients, sufficiently small relative perturbations of all the couplings preserve the regularity conditions, and so do small deformations of any finite number of couplings in general. We have also shown that theories admitting RBHs are dense, and therefore any theory yielding a singular solution may be completed into a theory with regular solutions in a non-unique way. The example given in~\eqref{eq:h_completed_rbh} is a particular representative case. This result has the important consequence that, from the effective field theory point of view, regularity cannot be ruled out at any order in curvature.
		
		Several questions remain open. It would be interesting to extend the analysis to more general quasi-topological theories, including non-polynomial~\cite{Borissova:2026wmn,Bueno:2025zaj} and non-local~\cite{Bueno:2026oyg} constructions, where the vacuum equations are no longer necessarily encoded by a single function. On the other hand, it would also be worth identifying a physical principle that naturally excludes deformations such as those introduced in~\eqref{eq:hN}, which allow for an unlimited number of horizons. 
		Extended definitions of energy conditions, or positivity of the entropy function, may provide such a restriction \cite{Yang:2021civ,Ye:2026lah}.
		
		It will also be interesting to study the thermodynamic properties of near-ultra-extremal configurations introduced above. In particular, understanding how the temperature, entropy and near-horizon scaling behave when four or more horizons approach one another may reveal whether these solutions admit a thermodynamic structure analogous to the extremal cases, or whether genuinely new behavior appears at higher degeneracy.

		\section*{Acknowledgments}
		We thank Giacomo Ageno and Anand Srinivasan for useful discussions. During the preparation of this manuscript, we used ChatGPT (Astra 6) and Claude (Opus 5.5) to assist with the development of ideas and consistency checks of selected equations. The authors take full responsibility for the final content and the validity of the results. PB was supported by
		a Proyecto de Consolidación Investigadora (CNS 2023-143822) from Spain’s Ministry of Science, Innovation and Universities, and by the grant PID2022-136224NB-C22, funded by MCIN/AEI/10.13039/501100011033/FEDER, UE. RAH is supported by a Willmore Fellowship at Durham University and by an STFC Ernest Rutherford Fellowship UKRI5636.
		MGM thanks the University of Barcelona for its hospitality during the preparation of this work.
		ÁJM is supported by the European Union’s Horizon Europe research and innovation programme under the Marie Sk\l odowska-Curie grant agreement No. 101202045 (RoBHin-ETG).

		\onecolumngrid
		\vspace{1cm}
		\begin{center}  
			{\Large\bf Appendices} 
		\end{center} 
		\appendix 
		
		\section{Quasi-topological densities to quintic order}\label{app:qt_densities}
		In this appendix we write in a compact and closed form the first five quasi-topological densities that can be written as invariants built from the curvature tensors. We will denote by $R$ the Ricci scalar, $Z_{ab} = R_{ab} - (1/D)g_{ab}R$ the traceless part of the Ricci tensor, and $W_{abcd}$ the Weyl tensor. The first five densities are then~\cite{Bueno:2019ycr,Bueno:2022res,Moreno:2023rfl}
		\begin{subequations}
			\begin{align}
				\mathcal Z_{(1)} &= R\;,
				\\[0.5em]
				\mathcal Z_{(2)} &= \frac{\mathcal Z_{(1)}^2}{D(D-1)} 
				- \frac{1}{D-2}
				\left[
				\frac{4Z_{ab}Z^{ab}}{D-2}
				- \frac{W_{abcd}W^{abcd}}{D-3}
				\right]\;,
				\\[0.5em]
				\mathcal Z_{(3)} &= \frac{3\mathcal Z_{(1)}\mathcal Z_{(2)}}{D(D-1)}
				- \frac{2\mathcal Z_{(1)}^3}{D^2(D-1)^2}
				+ \frac{24}{(D-2)^2(D-3)}
				\left[
				\frac{W_{ac}{}^{bd}Z_b^a Z_d^c}{D-2}
				- \frac{W_{acde}W^{bcde}Z_b^a}{D-4}
				\right] \notag \\
				&+ \frac{16 Z_a^b Z_b^c Z_c^a}{(D-2)^4}
				+ \frac{2(2D-3)W_{ab}{}^{cd}W_{cd}{}^{ef}W_{ef}{}^{ab}}
				{(D-2)(D-3)(D^3-9D^2+26D-22)}\;,
				\\[0.5em]
				\mathcal Z_{(4)} &= \frac{4\mathcal Z_{(1)}\mathcal Z_{(3)}-3\mathcal Z_{(2)}^2}{D(D-1)}
				+ \frac{96}{(D-2)^2(D-3)}
				\Bigg[
				\frac{(D-1)(W_{abcd}W^{abcd})^2}{8D(D-2)^2(D-3)}
				- \frac{4Z_{ac}Z_{de}W^{bdce}Z_b^a}{(D-2)^2(D-4)} \notag \\
				& - \frac{(2D-3)Z_e^f Z_f^e W_{abcd}W^{abcd}}
				{4(D-1)(D-2)^2}
				- \frac{2W_{acbd}W^{cefg}W^d{}_{efg}Z^{ab}}
				{D(D-3)(D-4)}
				+ \frac{(D^2-3D+3)(Z_b^a Z_a^b)^2}{D(D-1)(D-2)^3} \notag \\
				& - \frac{Z_b^a Z_c^b Z_d^c Z_a^d}{(D-2)^3}
				+ \frac{(2D-1)W_{abcd}W^{aecf}Z^{bd}Z_{ef}}
				{D(D-2)(D-3)}
				\Bigg]\;,
				\\[0.5em]
				\mathcal Z_{(5)} &= \frac{5\mathcal Z_{(1)}\mathcal Z_{(4)}-2\mathcal Z_{(2)}\mathcal Z_{(3)}}{D(D-1)}
				+ \frac{6\mathcal Z_{(1)}\mathcal Z_{(2)}^2-8\mathcal Z_{(1)}^2\mathcal Z_{(3)}}{D^2(D-1)^2}
				+ \frac{768 Z_b^a Z_c^b Z_d^c Z_e^d Z_a^e}{5(D-2)^6(D-3)(D-4)} \notag \\
				& + \frac{24(D-1)W_{ghij}W^{ghij}W_{ab}{}^{cd}W_{cd}{}^{ef}W_{ef}{}^{ab}}
				{D(D-2)^3(D-3)^2(D^3-9D^2+26D-22)} 
				- \frac{96(3D-1)W^{ghij}W_{ghij}W_{acde}W^{bcde}Z_b^a}
				{10D(D-1)(D-2)^4(D-3)^2(D-4)} \notag \\
				& - \frac{768(D^2-2D+2)Z_b^a Z_a^b Z_d^c Z_e^d Z_c^e}
				{D(D-1)^2(D-2)^6(D-3)(D-4)}
				- \frac{96(3D-1)(D^2+2D-4)W^{ghij}W_{ghij}Z_d^c Z_e^d Z_c^e}
				{D(D-1)(D+1)(D-2)^6(D-3)(D-4)} \notag \\
				& + \frac{24(15D^5-148D^4+527D^3-800D^2+472D-88)
					W_{ab}{}^{cd}W_{cd}{}^{ef}W_{ef}{}^{ab}Z_g^h Z_h^g}
				{D(D-1)(D-2)^3(D-3)(D-4)(D^5-15D^4+91D^3-277D^2+418D-242)} \notag \\
				& + \frac{960(D-1)}{(D-2)^4(D-3)^2}
				\Bigg[
				\frac{(5D^2-7D+6)Z_g^h Z_h^g W_{abcd}Z^{ac}Z^{bd}}
				{10D(D-1)^2(D-2)}
				- \frac{Z_b^a Z_c^b Z_d^c Z_{ef}W^{eafd}}
				{(D-1)(D-2)} \notag \\
				&- \frac{2(3D-1)Z^{ab}W_{acbd}Z^{ef}W_e{}^c{}_f{}^gZ_g^d}
				{D(D^2-1)(D-4)}
				+ \frac{(D-3)W_{acde}W^{bcde}Z_b^a Z_f^g Z_g^f}
				{5D(D-1)^2(D-4)} \notag \\
				&- \frac{(D-2)(D-3)(3D-2)Z_b^a Z_c^b W_{daef}W^{efgh}W_{gh}{}^{dc}}
				{4(D-1)^2(D-4)(D^2-6D+11)}
				+ \frac{W_{ghij}W^{ghij}Z^{ac}Z^{bd}W_{abcd}}
				{20D(D-1)^2}
				\Bigg]\;.
			\end{align}
		\end{subequations}
		The first density corresponds to the well-known Einstein-Hilbert term, while the second one is proportional to the usual Gauss-Bonnet term $\mathcal G = R^2-4R_{\mu\nu}R^{\mu\nu} + R^{\alpha\beta\mu\nu}R_{\alpha\beta\mu\nu}$ (though written in a non-standard form), coinciding with Lovelock theory at this stage. The proportionality factor is precisely
		\begin{equation}
			\mathcal Z_2 = \frac{\mathcal G}{(D-2)(D-3)}\,.	
		\end{equation} 
		The rest of quasi-topological densities beyond fifth order can be computed via the following recursive relation~\cite{Bueno:2019ycr},
		\begin{equation}
			\mathcal Z_{n+5} = \frac{3(n+3)}{(n+1)D(D-1)}\mathcal Z_1\mathcal Z_{n+4} - \frac{3(n+4)}{nD(D-1)}\mathcal Z_2\mathcal Z_{n+3} + \frac{(n+3)(n+4)}{n(n+1)D(D-1)}\mathcal Z_3\mathcal Z_{n+2}\,,\quad \forall n\geq1\,.
		\end{equation}
		With this prescription, theories of quasi-topological gravity of arbitrarily high curvature order can be naturally constructed. 	
		
		It is important to mention that quasi-topological gravities are not unique. In fact, one can consider curvature invariants $\mathcal T_n$ of order $n$ that, when evaluated on a spherically symmetric background, do not contribute to the equations of motion (see~\cite{Moreno:2023rfl}). In this case, the densities $\tilde{\mathcal Z}_n = \mathcal Z_n + \beta_n\mathcal T_n$ will also correspond to a quasi-topological theory with exactly the same equations of motion and dynamics on spherically symmetric backgrounds. This redundancy will play a role when considering non-spherically symmetric backgrounds, though for our purposes it is not relevant.	
		
		\section{Spherically symmetric solutions of the degenerate branch}\label{app:degenerate}
		In this appendix, we will review some representative examples of spherically symmetric spacetimes, regular as well as singular, that can be obtained as solutions of generic QT theories at the degenerate branch, defined by the condition that $h(\psi)=h'(\psi)=0$ at some particular value of $\psi$.
		
		\subsection{Einstein's static Universe}
		Consider the subset of spacetimes~\eqref{eq:ds2_deg} that are static, \(A(r,t)=A(r)\), with \(A(r)\equiv e^{2\Phi(r)}>0\) throughout the physical domain. In the particular case in which the redshift function is constant, \(\Phi(r)=\Phi_0\), a suitable rescaling of the time coordinate allows us to set \(g_{tt}=-1\). Assuming \(\psi_0>0\), we introduce the coordinate $\chi$ as
		\begin{equation}
			\chi=\arcsin\left(\sqrt{\psi_0}\,r\right)\,,
		\end{equation}
		in terms of which the metric takes the form
		\begin{equation}
			\dd s^2=-\dd t^2+\frac{1}{\psi_0}
			\left(\dd\chi^2+\sin^2\chi\dd\Omega_{D-2}^2\right).
		\end{equation}
		The radial coordinate \(r\) covers the range \(0\leq r\leq 1/\sqrt{\psi_0}\), corresponding initially to \(0\leq\chi\leq\pi/2\). However, the geometry is regular at \(\chi=\pi/2\), where the areal radius reaches its maximum value \(r=1/\sqrt{\psi_0}\), and it can be smoothly extended to \(0\leq\chi\leq\pi\). The resulting spatial sections are round \((D-1)\)-spheres of radius \(1/\sqrt{\psi_0}\), so the spacetime corresponds to the \(D\)-dimensional generalization of Einstein's static Universe~\cite{Einstein:1917ce,ORaifeartaigh:2017uct}.
		Notice, however, that here this geometry arises as a purely vacuum solution of the theory, whereas Einstein's original construction in GR required a homogeneous matter distribution and a cosmological constant.
		
		More generally, a non-constant redshift function \(\Phi(r)\) preserves the geometry of the spatial sections while introducing a non-trivial gravitational redshift. In terms of the coordinate \(\chi\), the metric can be written as
		\begin{equation}
			\dd s^2=-e^{2\Phi(\chi)}\dd t^2
			+\frac{1}{\psi_0}\left(\dd\chi^2+\sin^2\chi\dd\Omega_{D-2}^2\right)\,.
			\label{eq:deg_redshifted_ESU}
		\end{equation}
		Thus, while every constant-time slice remains a round \((D-1)\)-sphere, the full spacetime is no longer the Einstein static Universe unless \(\Phi\) is constant. Instead, \(\Phi(\chi)\) determines a position-dependent redshift along the spatial sphere.
		
		Provided \(A(r)=e^{2\Phi(r)}\) remains finite and strictly positive, no Killing horizon is generated. In particular, the surface $r=1/\sqrt{\psi_0}$, or equivalently $\chi=\pi/2$, is a regular surface at which the areal radius reaches a maximum. It therefore corresponds to the equator of the spatial sphere, rather than to a horizon or a wormhole throat. The latter would instead require a local minimum of the areal radius.
		
		Let us explicitly show that this degenerate branch cannot describe a traversable wormhole~\cite{Morris:1988cz}. Indeed, introducing the proper radial distance \(\ell\) on a constant-time slice,
		\begin{equation}
			\dd\ell=\frac{\dd r}{\sqrt{1-\psi_0r^2}},
		\end{equation}
		the areal radius satisfies
		\begin{equation}
			\left(\frac{\dd r}{\dd\ell}\right)^2=1-\psi_0r^2,
			\qquad
			\frac{\dd^2r}{\dd\ell^2}=-\psi_0r\,.
		\end{equation}
		A wormhole throat would require a local minimum of the areal radius, namely \(\dd r/\dd\ell=0\) and \(\dd^2r/\dd\ell^2>0\). For \(\psi_0>0\), the only stationary point occurs at \(r=1/\sqrt{\psi_0}\), where instead
		\begin{equation}
			\left.\frac{\dd^2r}{\dd\ell^2}\right|_{r=1/\sqrt{\psi_0}}
			=-\sqrt{\psi_0}<0,
		\end{equation}
		showing that this surface is a maximum of the areal radius. For \(\psi_0\leq0\), no stationary point exists at finite \(r\). Therefore, independently of the choice of the redshift function \(\Phi(r)\), the degenerate branch \(\psi=\psi_0\) does not contain traversable wormhole geometries.\footnote{This conclusion relies on the spherical topology of the transverse sections. For hyperbolic sections, the degenerate branch contains the static vacuum wormholes of~\cite{Dotti:2006cp,Dotti:2007az}.}
		
		Moreover, traversable wormholes are also excluded from the non-degenerate branch. Indeed, in that case \(g_{tt}g_{rr}=-1\), so the throat condition \(g^{rr}=0\) necessarily implies \(g_{tt}=0\), and the corresponding surface is a horizon rather than a traversable throat. Together with the previous result, this shows that traversable wormholes are absent from the static, spherically symmetric vacuum sector considered here.\footnote{
			Solutions closely related to traversable wormholes, usually referred to as black bounces, have been constructed explicitly in~\cite{Ling:2025ncw}. In those cases, however, the characteristic function is not globally well-defined and the construction requires patching different branches of the theory across a shell of constant radius. The physical interpretation of such constructions is not clear, so we will not consider them in the present work. Certain generalizations of these QT theories, however, can admit traversable wormholes or black-bounce geometries without requiring such fine-tuning~\cite{Mazza:2026taj}.
		}
		
		\subsection{Time-dependent vacua}
		The most interesting consequence of the degeneracy is that it allows for spherically symmetric, genuinely time-dependent vacuum solutions. Notice that, if \(A(r,t)=a(t)b(r)\) is separable, the time-dependent factor \(a(t)\) can be absorbed into a redefinition of the time coordinate, and the resulting spacetime is therefore static. On the other hand, for a genuinely non-separable function \(A(r,t)\), the time dependence cannot be removed in this way, giving rise to spherically symmetric vacuum geometries that are not static. The degenerate branch therefore provides an explicit violation of Birkhoff's theorem, as noted in \cite{Bueno:2025qjk},  analogous to the one known to occur at the degenerate point of EGB gravity (see \textit{e.g.,}~\cite{Dotti:2007az}).

		\subsection{Maximally symmetric spacetimes}
		Clearly, choosing \(A(r)=1-\psi_0 r^2\) yields a maximally symmetric spacetime: de Sitter for \(\psi_0>0\) and anti-de Sitter for \(\psi_0<0\). Minkowski spacetime is recovered only formally in the limit \(\psi_0\to0\), since the exact value \(\psi_0=0\) is excluded within the degenerate branch for theories admitting a well-defined Einstein limit.

		Since these geometries are simply the maximally symmetric vacua of the theory (at the degenerate point, $\psi_0$ is a double root of $h$), they do not provide any genuinely new features of the degenerate sector and we will not discuss them further.
		
		More generally, however, we can consider
		\begin{equation}
			A(r)=e^{2\Phi(r)}\left(1-\psi_0r^2\right),
		\end{equation}
		where \(\Phi(r)\) is an arbitrary regular function. In this case,
		\begin{equation}
			g_{tt}g_{rr}=-e^{2\Phi(r)}\neq -1,
		\end{equation}
		and the spacetime is no longer maximally symmetric for a generic choice of \(\Phi(r)\). Nevertheless, for \(\psi_0>0\) and finite \(\Phi(r)\), the surface \(r=1/\sqrt{\psi_0}\) remains a Killing horizon. Therefore, the degenerate branch contains an infinite family of inequivalent redshift deformations of the de Sitter static patch, sharing the same radial geometry and horizon location while differing in their redshift profile and spacetime curvature.
		
		In this sense, the arbitrary function \(\Phi(r)\) can be interpreted as a form of functional gravitational hair associated with the degeneracy of the field equations. Unlike the usual notion of black hole hair, however, this freedom is not characterized by a finite number of additional charges, but by an arbitrary function left unconstrained by the equations of motion.

		\twocolumngrid

		\bibliographystyle{JHEP}
		\bibliography{refs1.bib}

@article{Penrose:1964wq,
	author = "Penrose, Roger",
	title = "{Gravitational collapse and space-time singularities}",
	doi = "10.1103/PhysRevLett.14.57",
	journal = "Phys. Rev. Lett.",
	volume = "14",
	pages = "57--59",
	year = "1965"
}

@article{Murcia:2026pmg,
	author = "Murcia, {\'A}ngel J.",
	title = "{All theories of gravity with second-order traced field equations}",
	eprint = "2610.12097",
	archivePrefix = "arXiv",
	primaryClass = "gr-qc",
	month = "10",
	year = "2026"
}

@article{Spanier:1963,
	author = "Spanier, Edwin",
	title = "{Quasi-topologies}",
	journal = "Duke Math. J.",
	volume = "30",
	number = "1",
	pages = "1--14",
	year = "1963",
	doi = "10.1215/S0012-7094-63-03001-1"
}

@article{Arrechea:2025nlq,
	author = "Arrechea, Julio and Liberati, Stefano and Neshat, Hooman and Vellucci, Vania",
	title = "{From de Sitter to anti{\textendash}de Sitter singularity regularization: Theory and phenomenology}",
	eprint = "2509.13421",
	archivePrefix = "arXiv",
	primaryClass = "gr-qc",
	reportNumber = "DJ14147",
	doi = "10.1103/919y-m4yj",
	journal = "Phys. Rev. D",
	volume = "112",
	number = "12",
	pages = "124029",
	year = "2025"
}

@article{Sakharov:1966aja,
	author = "Sakharov, A. D.",
	title = "{Nachal'naia stadija rasshirenija Vselennoj i vozniknovenije neodnorodnosti raspredelenija veshchestva}",
	journal = "Sov. Phys. JETP",
	volume = "22",
	pages = "241",
	year = "1966"
}

@article{Poisson:1988wc,
	author = "Poisson, Eric and Israel, W.",
	title = "{Structure of the Black Hole Nucleus}",
	doi = "10.1088/0264-9381/5/12/002",
	journal = "Class. Quant. Grav.",
	volume = "5",
	pages = "L201--L205",
	year = "1988"
}

@article{Carballo-Rubio:2018pmi,
	author = "Carballo-Rubio, Ra{\'u}l and Di Filippo, Francesco and Liberati, Stefano and Pacilio, Costantino and Visser, Matt",
	title = "{On the viability of regular black holes}",
	eprint = "1805.02675",
	archivePrefix = "arXiv",
	primaryClass = "gr-qc",
	doi = "10.1007/JHEP07(2018)023",
	journal = "JHEP",
	volume = "07",
	pages = "023",
	year = "2018"
}

@article{Carballo-Rubio:2026gwg,
	author = "Carballo-Rubio, Ra{\'u}l and Di Filippo, Francesco and Liberati, Stefano and Visser, Matt",
	title = "{Semiclassical regularity of compact trapped regions: From dynamical horizons to inner extremality}",
	eprint = "2607.03916",
	archivePrefix = "arXiv",
	primaryClass = "gr-qc",
	month = "7",
	year = "2026"
}

@article{Carballo-Rubio:2025fnc,
	author = "Carballo-Rubio, Ra{\'u}l and others",
	title = "{Towards a non-singular paradigm of black hole physics}",
	eprint = "2501.05505",
	archivePrefix = "arXiv",
	primaryClass = "gr-qc",
	doi = "10.1088/1475-7516/2025/05/003",
	journal = "JCAP",
	volume = "05",
	pages = "003",
	year = "2025"
}

@article{Carballo-Rubio:2022kad,
	author = "Carballo-Rubio, Ra\'ul and Di Filippo, Francesco and Liberati, Stefano and Pacilio, Costantino and Visser, Matt",
	title = "{Regular black holes without mass inflation instability}",
	eprint = "2205.13556",
	archivePrefix = "arXiv",
	primaryClass = "gr-qc",
	reportNumber = "YITP-22-53",
	doi = "10.1007/JHEP09(2022)118",
	journal = "JHEP",
	volume = "09",
	pages = "118",
	year = "2022"
}

@article{Cano:2020qhy,
	author = "Cano, Pablo A. and Murcia, {\'A}ngel",
	title = "{Electromagnetic Quasitopological Gravities}",
	eprint = "2007.04331",
	archivePrefix = "arXiv",
	primaryClass = "hep-th",
	doi = "10.1007/JHEP10(2020)125",
	journal = "JHEP",
	volume = "10",
	pages = "125",
	year = "2020"
}

@article{Cano:2020ezi,
	author = "Cano, Pablo A. and Murcia, {\'A}ngel",
	title = {{Resolution of Reissner-Nordstr{\"o}m singularities by higher-derivative corrections}},
	eprint = "2006.15149",
	archivePrefix = "arXiv",
	primaryClass = "hep-th",
	reportNumber = "IFT-UAM/CSIC-20-99",
	doi = "10.1088/1361-6382/abd923",
	journal = "Class. Quant. Grav.",
	volume = "38",
	number = "7",
	pages = "075014",
	year = "2021"
}

@article{Bueno:2021krl,
	author = "Bueno, Pablo and Cano, Pablo A. and Moreno, Javier and van der Velde, Guido",
	title = "{Regular black holes in three dimensions}",
	eprint = "2104.10172",
	archivePrefix = "arXiv",
	primaryClass = "gr-qc",
	doi = "10.1103/PhysRevD.104.L021501",
	journal = "Phys. Rev. D",
	volume = "104",
	number = "2",
	pages = "L021501",
	year = "2021"
}

@article{Cano:2022ord,
	author = "Cano, Pablo A. and Murcia, {\'A}ngel J. and Rivadulla S{\'a}nchez, Alberto and Zhang, Xuao",
	title = "{Higher-derivative holography with a chemical potential}",
	eprint = "2202.10473",
	archivePrefix = "arXiv",
	primaryClass = "hep-th",
	reportNumber = "IFT-UAM/CSIC-22-20",
	doi = "10.1007/JHEP07(2022)010",
	journal = "JHEP",
	volume = "07",
	pages = "010",
	year = "2022"
}

@article{Bueno:2025qjk,
	author = "Bueno, Pablo and Hennigar, Robie A. and Murcia, {\'A}ngel J.",
	title = "{Birkhoff implies quasi-topological}",
	eprint = "2510.25823",
	archivePrefix = "arXiv",
	primaryClass = "gr-qc",
	doi = "10.1088/1361-6382/ae6600",
	journal = "Class. Quant. Grav.",
	volume = "43",
	number = "9",
	pages = "095020",
	year = "2026"
}

@article{Hawking:1970zqf,
	author = "Hawking, S. W. and Penrose, R.",
	title = "{The Singularities of gravitational collapse and cosmology}",
	doi = "10.1098/rspa.1970.0021",
	journal = "Proc. Roy. Soc. Lond. A",
	volume = "314",
	pages = "529--548",
	year = "1970"
}

@book{Hawking:1973uf,
	author = "Hawking, Stephen W. and Ellis, George F. R.",
	title = "{The Large Scale Structure of Space-Time}",
	doi = "10.1017/9781009253161",
	isbn = "978-1-009-25316-1, 978-1-009-25315-4, 978-0-521-20016-5, 978-0-521-09906-6, 978-0-511-82630-6, 978-0-521-09906-6",
	publisher = "Cambridge University Press",
	series = "Cambridge Monographs on Mathematical Physics",
	month = "2",
	year = "2023"
}

@article{Padmanabhan:2011ex,
	author = "Padmanabhan, T.",
	title = "{Some aspects of field equations in generalised theories of gravity}",
	eprint = "1109.3846",
	archivePrefix = "arXiv",
	primaryClass = "gr-qc",
	doi = "10.1103/PhysRevD.84.124041",
	journal = "Phys. Rev. D",
	volume = "84",
	pages = "124041",
	year = "2011"
}

@article{Lovelock:1970zsf,
	author = "Lovelock, David",
	title = "{Divergence-free tensorial concomitants}",
	doi = "10.1007/BF01817753",
	journal = "Aequat. Math.",
	volume = "4",
	number = "1",
	pages = "127--138",
	year = "1970"
}

@article{Lovelock:1971yv,
	author = "Lovelock, D.",
	title = "{The Einstein tensor and its generalizations}",
	doi = "10.1063/1.1665613",
	journal = "J. Math. Phys.",
	volume = "12",
	pages = "498--501",
	year = "1971"
}

@article{Oliva:2010eb,
	author = "Oliva, Julio and Ray, Sourya",
	title = "{A new cubic theory of gravity in five dimensions: Black hole, Birkhoff's theorem and C-function}",
	eprint = "1003.4773",
	archivePrefix = "arXiv",
	primaryClass = "gr-qc",
	reportNumber = "CECS-PHY-10-03",
	doi = "10.1088/0264-9381/27/22/225002",
	journal = "Class. Quant. Grav.",
	volume = "27",
	pages = "225002",
	year = "2010"
}

@article{Myers:2010ru,
	author = "Myers, Robert C. and Robinson, Brandon",
	title = "{Black Holes in Quasi-topological Gravity}",
	eprint = "1003.5357",
	archivePrefix = "arXiv",
	primaryClass = "gr-qc",
	doi = "10.1007/JHEP08(2010)067",
	journal = "JHEP",
	volume = "08",
	pages = "067",
	year = "2010"
}

@article{Dehghani:2011vu,
	author = "Dehghani, M. H. and Bazrafshan, A. and Mann, R. B. and Mehdizadeh, M. R. and Ghanaatian, M. and Vahidinia, M. H.",
	title = "{Black Holes in Quartic Quasitopological Gravity}",
	eprint = "1109.4708",
	archivePrefix = "arXiv",
	primaryClass = "hep-th",
	doi = "10.1103/PhysRevD.85.104009",
	journal = "Phys. Rev. D",
	volume = "85",
	pages = "104009",
	year = "2012"
}

@article{Ahmed:2017jod,
	author = "Ahmed, Jamil and Hennigar, Robie A. and Mann, Robert B. and Mir, Mozhgan",
	title = "{Quintessential Quartic Quasi-topological Quartet}",
	eprint = "1703.11007",
	archivePrefix = "arXiv",
	primaryClass = "hep-th",
	doi = "10.1007/JHEP05(2017)134",
	journal = "JHEP",
	volume = "05",
	pages = "134",
	year = "2017"
}

@article{Cisterna:2017umf,
	author = "Cisterna, Adolfo and Guajardo, Luis and Hassaine, Mokhtar and Oliva, Julio",
	title = "{Quintic quasi-topological gravity}",
	eprint = "1702.04676",
	archivePrefix = "arXiv",
	primaryClass = "hep-th",
	doi = "10.1007/JHEP04(2017)066",
	journal = "JHEP",
	volume = "04",
	pages = "066",
	year = "2017"
}

@article{Bueno:2019ycr,
	author = "Bueno, Pablo and Cano, Pablo A. and Hennigar, Robie A.",
	title = "{(Generalized) quasi-topological gravities at all orders}",
	eprint = "1909.07983",
	archivePrefix = "arXiv",
	primaryClass = "hep-th",
	reportNumber = "IFT-UAM/CSIC-19-124",
	doi = "10.1088/1361-6382/ab5410",
	journal = "Class. Quant. Grav.",
	volume = "37",
	number = "1",
	pages = "015002",
	year = "2020"
}

@article{Bueno:2022res,
	author = "Bueno, Pablo and Cano, Pablo A. and Hennigar, Robie A. and Lu, Mengqi and Moreno, Javier",
	title = "{Generalized quasi-topological gravities: the whole shebang}",
	eprint = "2203.05589",
	archivePrefix = "arXiv",
	primaryClass = "hep-th",
	reportNumber = "CERN-TH-2022-038",
	doi = "10.1088/1361-6382/aca236",
	journal = "Class. Quant. Grav.",
	volume = "40",
	number = "1",
	pages = "015004",
	year = "2023"
}

@article{Moreno:2023rfl,
	author = "Moreno, Javier and Murcia, {\'A}ngel J.",
	title = "{Classification of generalized quasitopological gravities}",
	eprint = "2304.08510",
	archivePrefix = "arXiv",
	primaryClass = "gr-qc",
	doi = "10.1103/PhysRevD.108.044016",
	journal = "Phys. Rev. D",
	volume = "108",
	number = "4",
	pages = "044016",
	year = "2023"
}

@article{Bueno:2024zsx,
	author = "Bueno, Pablo and Cano, Pablo A. and Hennigar, Robie A. and Murcia, {\'A}ngel J.",
	title = "{Regular black holes from thin-shell collapse}",
	eprint = "2412.02740",
	archivePrefix = "arXiv",
	primaryClass = "gr-qc",
	doi = "10.1103/PhysRevD.111.104009",
	journal = "Phys. Rev. D",
	volume = "111",
	number = "10",
	pages = "104009",
	year = "2025"
}

@article{Bueno:2024eig,
	author = "Bueno, Pablo and Cano, Pablo A. and Hennigar, Robie A. and Murcia, {\'A}ngel J.",
	title = "{Dynamical Formation of Regular Black Holes}",
	eprint = "2412.02742",
	archivePrefix = "arXiv",
	primaryClass = "gr-qc",
	doi = "10.1103/PhysRevLett.134.181401",
	journal = "Phys. Rev. Lett.",
	volume = "134",
	number = "18",
	pages = "181401",
	year = "2025"
}

@article{Horndeski:1974wa,
	author = "Horndeski, Gregory Walter",
	title = "{Second-order scalar-tensor field equations in a four-dimensional space}",
	doi = "10.1007/BF01807638",
	journal = "Int. J. Theor. Phys.",
	volume = "10",
	pages = "363--384",
	year = "1974"
}

@article{Bueno:2024dgm,
	author = "Bueno, Pablo and Cano, Pablo A. and Hennigar, Robie A.",
	title = "{Regular black holes from pure gravity}",
	eprint = "2403.04827",
	archivePrefix = "arXiv",
	primaryClass = "gr-qc",
	doi = "10.1016/j.physletb.2025.139260",
	journal = "Phys. Lett. B",
	volume = "861",
	pages = "139260",
	year = "2025"
}

@article{Hayward:2005gi,
	author = "Hayward, Sean A.",
	title = "{Formation and evaporation of regular black holes}",
	eprint = "gr-qc/0506126",
	archivePrefix = "arXiv",
	doi = "10.1103/PhysRevLett.96.031103",
	journal = "Phys. Rev. Lett.",
	volume = "96",
	pages = "031103",
	year = "2006"
}

@INPROCEEDINGS{Bardeen1968,
	author = {{Bardeen}, James},
	title = "{Non-singular general relativistic gravitational collapse}",
	booktitle = {Proceedings of the 5th International Conference on Gravitation and the Theory of Relativity},
	year = 1968,
	month = sep,
	pages = {87},
	adsurl = {https://ui.adsabs.harvard.edu/abs/1968qtr..conf...87B}
}

@ARTICLE{Dymnikova:1992,
	author = {{Dymnikova}, Irina},
	title = "{Vacuum nonsingular black hole}",
	journal = {General Relativity and Gravitation},
	year = 1992,
	month = mar,
	volume = {24},
	number = {3},
	pages = {235-242},
	doi = {10.1007/BF00760226},
	adsurl = {https://ui.adsabs.harvard.edu/abs/1992GReGr..24..235D}
}

@article{Bueno:2025gjg,
	author = "Bueno, Pablo and Cano, Pablo A. and Hennigar, Robie A. and Murcia, {\'A}ngel J. and Vicente-Cano, Aitor",
	title = "{Regular black holes from Oppenheimer-Snyder collapse}",
	eprint = "2505.09680",
	archivePrefix = "arXiv",
	primaryClass = "gr-qc",
	doi = "10.1103/qrbb-mdvm",
	journal = "Phys. Rev. D",
	volume = "112",
	number = "6",
	pages = "064039",
	year = "2025"
}

@article{Bueno:2026wjt,
	author = "Bueno, Pablo and Cano, Pablo A. and Hennigar, Robie A. and Murcia, {\'A}ngel J. and Vicente-Cano, Aitor",
	title = "{Thin shells in Lovelock and Quasi-topological gravities}",
	eprint = "2610.18977",
	archivePrefix = "arXiv",
	primaryClass = "gr-qc",
	month = "10",
	year = "2026"
}

@article{DiFilippo:2024mwm,
	author = "Di Filippo, Francesco and Kol{\'a}{\v{r}}, Ivan and Kubiznak, David",
	title = "{Inner-extremal regular black holes from pure gravity}",
	eprint = "2404.07058",
	archivePrefix = "arXiv",
	primaryClass = "gr-qc",
	doi = "10.1103/PhysRevD.111.L041505",
	journal = "Phys. Rev. D",
	volume = "111",
	number = "4",
	pages = "L041505",
	year = "2025"
}

@article{Frolov:2026rcm,
	author = "Frolov, Valeri P. and Zelnikov, Andrei",
	title = "{Regular black holes in quasitopological gravity: Null shells and mass inflation}",
	eprint = "2601.01861",
	archivePrefix = "arXiv",
	primaryClass = "gr-qc",
	reportNumber = "Alberta Thy 5-25",
	doi = "10.1103/nrd7-j6hs",
	journal = "Phys. Rev. D",
	volume = "113",
	number = "8",
	pages = "084007",
	year = "2026"
}

@article{Bueno:2025zaj,
	author = "Bueno, Pablo and Cano, Pablo A. and Hennigar, Robie A. and Murcia, {\'A}ngel J.",
	title = "{Regular black hole formation in four-dimensional nonpolynomial gravities}",
	eprint = "2509.19016",
	archivePrefix = "arXiv",
	primaryClass = "gr-qc",
	doi = "10.1103/8f3j-zcxh",
	journal = "Phys. Rev. D",
	volume = "113",
	number = "2",
	pages = "024019",
	year = "2026"
}

@article{Borissova:2026wmn,
	author = "Borissova, Johanna and Carballo-Rubio, Ra{\'u}l",
	title = "{Regular black holes from pure gravity in four dimensions}",
	eprint = "2602.16773",
	archivePrefix = "arXiv",
	primaryClass = "gr-qc",
	reportNumber = "Imperial/TP/2026/JB/02",
	doi = "10.1103/6x2z-qbkh",
	journal = "Phys. Rev. D",
	volume = "113",
	number = "12",
	pages = "124004",
	year = "2026"
}

@article{Konoplya:2024hfg,
	author = "Konoplya, R. A. and Zhidenko, A.",
	title = "{Infinite tower of higher-curvature corrections: Quasinormal modes and late-time behavior of D-dimensional regular black holes}",
	eprint = "2403.07848",
	archivePrefix = "arXiv",
	primaryClass = "gr-qc",
	doi = "10.1103/PhysRevD.109.104005",
	journal = "Phys. Rev. D",
	volume = "109",
	number = "10",
	pages = "104005",
	year = "2024"
}

@article{Konoplya:2024kih,
	author = "Konoplya, R. A. and Zhidenko, A.",
	title = "{Dymnikova black hole from an infinite tower of higher-curvature corrections}",
	eprint = "2404.09063",
	archivePrefix = "arXiv",
	primaryClass = "gr-qc",
	doi = "10.1016/j.physletb.2024.138945",
	journal = "Phys. Lett. B",
	volume = "856",
	pages = "138945",
	year = "2024"
}

@article{Ma:2024olw,
	author = "Ma, Tian-Xiang and Fang, Tie-Feng and Wang, Yong-Qiang",
	title = "{Boson stars and their frozen states in an infinite tower of higher-derivative gravity}",
	eprint = "2406.08813",
	archivePrefix = "arXiv",
	primaryClass = "gr-qc",
	doi = "10.1140/epjc/s10052-025-14252-4",
	journal = "Eur. Phys. J. C",
	volume = "85",
	number = "5",
	pages = "542",
	year = "2025"
}

@article{Frolov:2024hhe,
	author = "Frolov, Valeri P. and Koek, Alex and Soto, Jose Pinedo and Zelnikov, Andrei",
	title = "{Regular black holes inspired by quasitopological gravity}",
	eprint = "2411.16050",
	archivePrefix = "arXiv",
	primaryClass = "gr-qc",
	reportNumber = "Alberta Thy 11-24",
	doi = "10.1103/PhysRevD.111.044034",
	journal = "Phys. Rev. D",
	volume = "111",
	number = "4",
	pages = "044034",
	year = "2025"
}

@article{Wang:2024zlq,
	author = "Wang, Sheng-Wei and Wu, Shan-Ping and Wei, Shao-Wen",
	title = "{Are regular black holes from pure gravity classified within the same thermodynamical topology?}",
	eprint = "2412.05811",
	archivePrefix = "arXiv",
	primaryClass = "gr-qc",
	doi = "10.1016/j.physletb.2025.139402",
	journal = "Phys. Lett. B",
	volume = "864",
	pages = "139402",
	year = "2025"
}

@article{Bueno:2026dln,
	author = "Bueno, Pablo and Hennigar, Robie A. and Murcia, {\'A}ngel J. and Vicente-Cano, Aitor",
	title = "{Regular geometries from singular matter in quasitopological gravity}",
	eprint = "2603.10110",
	archivePrefix = "arXiv",
	primaryClass = "gr-qc",
	doi = "10.1103/3mk3-j8vs",
	journal = "Phys. Rev. D",
	volume = "114",
	number = "2",
	pages = "024070",
	year = "2026"
}

@article{Tsuda:2026xjc,
	author = "Tsuda, Ren and Suzuki, Ryotaku and Tomizawa, Shinya",
	title = "{Fan-Wang type regular black holes in quasitopological gravity}",
	eprint = "2602.16754",
	archivePrefix = "arXiv",
	primaryClass = "gr-qc",
	reportNumber = "TTI-MATHPHYS-39",
	doi = "10.1103/q4kl-6d3z",
	journal = "Phys. Rev. D",
	volume = "114",
	number = "2",
	pages = "024001",
	year = "2026"
}

@article{Bueno:2026oyg,
	author = "Bueno, Pablo and Cano, Pablo A. and Hennigar, Robie A. and Murcia, {\'A}ngel J.",
	title = "{Regular Black Holes in Nonlocal Quasitopological Gravity}",
	eprint = "2607.07790",
	archivePrefix = "arXiv",
	primaryClass = "gr-qc",
	month = "7",
	year = "2026"
}

@article{Hennigar:2025yqm,
	author = "Hennigar, Robie A. and Kubiz{\v{n}}{\'a}k, David and Murk, Sebastian and Soranidis, Ioannis",
	title = "{Thermodynamics of regular black holes in anti-de Sitter space}",
	eprint = "2505.11623",
	archivePrefix = "arXiv",
	primaryClass = "gr-qc",
	doi = "10.1007/JHEP11(2025)121",
	journal = "JHEP",
	volume = "11",
	pages = "121",
	year = "2025"
}

@article{Ayon-Beato:1998hmi,
	author = "Ayon-Beato, Eloy and Garcia, Alberto",
	title = "{Regular black hole in general relativity coupled to nonlinear electrodynamics}",
	eprint = "gr-qc/9911046",
	archivePrefix = "arXiv",
	doi = "10.1103/PhysRevLett.80.5056",
	journal = "Phys. Rev. Lett.",
	volume = "80",
	pages = "5056--5059",
	year = "1998"
}

@article{Bronnikov:2000vy,
	author = "Bronnikov, Kirill A.",
	title = "{Regular magnetic black holes and monopoles from nonlinear electrodynamics}",
	eprint = "gr-qc/0006014",
	archivePrefix = "arXiv",
	doi = "10.1103/PhysRevD.63.044005",
	journal = "Phys. Rev. D",
	volume = "63",
	pages = "044005",
	year = "2001"
}

@article{Ayon-Beato:2000mjt,
	author = "Ayon-Beato, Eloy and Garcia, Alberto",
	title = "{The Bardeen model as a nonlinear magnetic monopole}",
	eprint = "gr-qc/0009077",
	archivePrefix = "arXiv",
	doi = "10.1016/S0370-2693(00)01125-4",
	journal = "Phys. Lett. B",
	volume = "493",
	pages = "149--152",
	year = "2000"
}

@article{Bronnikov:2000yz,
	author = "Bronnikov, K. A.",
	title = "{Comment on `Regular black hole in general relativity coupled to nonlinear electrodynamics'}",
	doi = "10.1103/PhysRevLett.85.4641",
	journal = "Phys. Rev. Lett.",
	volume = "85",
	pages = "4641",
	year = "2000"
}

@article{Ayon-Beato:2004ywd,
	author = "Ayon-Beato, Eloy and Garcia, Alberto",
	title = "{Four parametric regular black hole solution}",
	eprint = "hep-th/0403229",
	archivePrefix = "arXiv",
	reportNumber = "CECS-PHY-04-06",
	doi = "10.1007/s10714-005-0050-y",
	journal = "Gen. Rel. Grav.",
	volume = "37",
	pages = "635",
	year = "2005"
}

@article{Dymnikova:2004zc,
	author = "Dymnikova, Irina",
	title = "{Regular electrically charged structures in nonlinear electrodynamics coupled to general relativity}",
	eprint = "gr-qc/0407072",
	archivePrefix = "arXiv",
	doi = "10.1088/0264-9381/21/18/009",
	journal = "Class. Quant. Grav.",
	volume = "21",
	pages = "4417--4429",
	year = "2004"
}

@article{Berej:2006cc,
	author = "Berej, Waldemar and Matyjasek, Jerzy and Tryniecki, Dariusz and Woronowicz, Mariusz",
	title = "{Regular black holes in quadratic gravity}",
	eprint = "hep-th/0606185",
	archivePrefix = "arXiv",
	doi = "10.1007/s10714-006-0270-9",
	journal = "Gen. Rel. Grav.",
	volume = "38",
	pages = "885--906",
	year = "2006"
}

@article{Balart:2014jia,
	author = "Balart, Leonardo and Vagenas, Elias C.",
	title = "{Regular black hole metrics and the weak energy condition}",
	eprint = "1401.2136",
	archivePrefix = "arXiv",
	primaryClass = "gr-qc",
	doi = "10.1016/j.physletb.2014.01.024",
	journal = "Phys. Lett. B",
	volume = "730",
	pages = "14--17",
	year = "2014"
}

@article{Fan:2016rih,
	author = "Fan, Zhong-Ying",
	title = "{Critical phenomena of regular black holes in anti-de Sitter space-time}",
	eprint = "1609.04489",
	archivePrefix = "arXiv",
	primaryClass = "hep-th",
	doi = "10.1140/epjc/s10052-017-4830-9",
	journal = "Eur. Phys. J. C",
	volume = "77",
	number = "4",
	pages = "266",
	year = "2017"
}

@article{Bronnikov:2017sgg,
	author = "Bronnikov, K. A.",
	title = "{Nonlinear electrodynamics, regular black holes and wormholes}",
	eprint = "1711.00087",
	archivePrefix = "arXiv",
	primaryClass = "gr-qc",
	doi = "10.1142/S0218271818410055",
	journal = "Int. J. Mod. Phys. D",
	volume = "27",
	number = "06",
	pages = "1841005",
	year = "2018"
}

@article{Murk:2024nod,
	author = "Murk, Sebastian and Soranidis, Ioannis",
	title = "{Light rings and causality for nonsingular ultracompact objects sourced by nonlinear electrodynamics}",
	eprint = "2406.07957",
	archivePrefix = "arXiv",
	primaryClass = "gr-qc",
	doi = "10.1103/PhysRevD.110.044064",
	journal = "Phys. Rev. D",
	volume = "110",
	number = "4",
	pages = "044064",
	year = "2024"
}

@article{PinedoSoto:2026hfm,
	author = "Pinedo Soto, Jose and Frolov, Valeri P.",
	title = "{Charged black holes in quasitopological gravity coupled to Born-Infeld nonlinear electrodynamics}",
	eprint = "2604.06632",
	archivePrefix = "arXiv",
	primaryClass = "gr-qc",
	doi = "10.1103/8px4-2pyk",
	journal = "Phys. Rev. D",
	volume = "113",
	number = "12",
	pages = "124044",
	year = "2026"
}

@article{Morris:1988cz,
	author = "Morris, M. S. and Thorne, K. S.",
	title = "{Wormholes in space-time and their use for interstellar travel: A tool for teaching general relativity}",
	doi = "10.1119/1.15620",
	journal = "Am. J. Phys.",
	volume = "56",
	pages = "395--412",
	year = "1988"
}

@article{Arnowitt:1960es,
	author = "Arnowitt, Richard L. and Deser, Stanley and Misner, Charles W.",
	title = "{Canonical variables for general relativity}",
	doi = "10.1103/PhysRev.117.1595",
	journal = "Phys. Rev.",
	volume = "117",
	pages = "1595--1602",
	year = "1960"
}

@article{Arnowitt:1962hi,
	author = "Arnowitt, Richard L. and Deser, Stanley and Misner, Charles W.",
	title = "{The Dynamics of general relativity}",
	eprint = "gr-qc/0405109",
	archivePrefix = "arXiv",
	doi = "10.1007/s10714-008-0661-1",
	journal = "Gen. Rel. Grav.",
	volume = "40",
	pages = "1997--2027",
	year = "2008"
}

@article{Chamseddine:1989nu,
	author = "Chamseddine, Ali H.",
	title = "{Topological Gauge Theory of Gravity in Five-dimensions and All Odd Dimensions}",
	reportNumber = "Print-89-0828 (ZURICH)",
	doi = "10.1016/0370-2693(89)91312-9",
	journal = "Phys. Lett. B",
	volume = "233",
	pages = "291--294",
	year = "1989"
}

@article{Anabalon:2024abz,
	author = "Anabalon, Andres and Astefanesei, Dumitru and Cisterna, Adolfo and Izaurieta, Fernando and Oliva, Julio and Quijada, Constanza and Quinzacara, Cristian",
	title = "{Rotating and accelerating AdS black holes in Einstein-Gauss-Bonnet gravity}",
	eprint = "2404.04691",
	archivePrefix = "arXiv",
	primaryClass = "hep-th",
	doi = "10.1016/j.physletb.2024.139000",
	journal = "Phys. Lett. B",
	volume = "857",
	pages = "139000",
	year = "2024"
}

@article{Anabalon:2009kq,
	author = "Anabalon, Andres and Deruelle, Nathalie and Morisawa, Yoshiyuki and Oliva, Julio and Sasaki, Misao and Tempo, David and Troncoso, Ricardo",
	title = "{Kerr-Schild ansatz in Einstein-Gauss-Bonnet gravity: An exact vacuum solution in five dimensions}",
	eprint = "0812.3194",
	archivePrefix = "arXiv",
	primaryClass = "hep-th",
	reportNumber = "CECS-PHY-08-21, AEI-2008-091, YITP-08-94",
	doi = "10.1088/0264-9381/26/6/065002",
	journal = "Class. Quant. Grav.",
	volume = "26",
	pages = "065002",
	year = "2009"
}

@article{Anabalon:2010ns,
	author = "Anabalon, Andres and Deruelle, Nathalie and Tempo, David and Troncoso, Ricardo",
	editor = "Calzetta, Esteban and Domenech, Graciela and Ferraro, Rafael and Giribet, Gaston and Laura, Roberto and Nunez, Carmen",
	title = "{Remarks on the Myers-Perry and Einstein Gauss-Bonnet Rotating Solutions}",
	eprint = "1009.3030",
	archivePrefix = "arXiv",
	primaryClass = "gr-qc",
	doi = "10.1142/S0218271811018974",
	journal = "Int. J. Mod. Phys. D",
	volume = "20",
	pages = "639--647",
	year = "2011"
}

@article{Camanho:2014apa,
	author = "Camanho, Xian O. and Edelstein, Jose D. and Maldacena, Juan and Zhiboedov, Alexander",
	title = "{Causality Constraints on Corrections to the Graviton Three-Point Coupling}",
	eprint = "1407.5597",
	archivePrefix = "arXiv",
	primaryClass = "hep-th",
	doi = "10.1007/JHEP02(2016)020",
	journal = "JHEP",
	volume = "02",
	pages = "020",
	year = "2016"
}

@article{Dotti:2006cp,
	author = "Dotti, Gustavo and Oliva, Julio and Troncoso, Ricardo",
	title = "{Static wormhole solution for higher-dimensional gravity in vacuum}",
	eprint = "hep-th/0607062",
	archivePrefix = "arXiv",
	reportNumber = "CECS-PHY-06-19",
	doi = "10.1103/PhysRevD.75.024002",
	journal = "Phys. Rev. D",
	volume = "75",
	pages = "024002",
	year = "2007"
}

@article{Dotti:2007az,
	author = "Dotti, Gustavo and Oliva, Julio and Troncoso, Ricardo",
	title = "{Exact solutions for the Einstein-Gauss-Bonnet theory in five dimensions: Black holes, wormholes and spacetime horns}",
	eprint = "0706.1830",
	archivePrefix = "arXiv",
	primaryClass = "hep-th",
	reportNumber = "CECS-PHY-07-09",
	doi = "10.1103/PhysRevD.76.064038",
	journal = "Phys. Rev. D",
	volume = "76",
	pages = "064038",
	year = "2007"
}

@article{Giribet:2014hpa,
	author = "Giribet, Gaston and Merino, Nelson and Miskovic, Olivera and Zanelli, Jorge",
	title = "{Black hole solutions in Chern-Simons AdS supergravity}",
	eprint = "1406.3096",
	archivePrefix = "arXiv",
	primaryClass = "hep-th",
	doi = "10.1007/JHEP08(2014)083",
	journal = "JHEP",
	volume = "08",
	pages = "083",
	year = "2014"
}

@article{Borissova:2026krh,
	author = "Borissova, Johanna",
	title = "{All 2D generalized dilaton theories from d{\ensuremath{\geq}}4 gravities}",
	eprint = "2603.06786",
	archivePrefix = "arXiv",
	primaryClass = "hep-th",
	reportNumber = "Imperial/TP/2026/JB/03",
	doi = "10.1103/v3j1-j6jg",
	journal = "Phys. Rev. D",
	volume = "113",
	number = "12",
	pages = "124088",
	year = "2026"
}

@article{Deser:2005pc,
	author = "Deser, Stanley and Ryzhov, A. V.",
	title = "{Curvature invariants of static spherically symmetric geometries}",
	eprint = "gr-qc/0505039",
	archivePrefix = "arXiv",
	reportNumber = "BRX-TH-563",
	doi = "10.1088/0264-9381/22/16/012",
	journal = "Class. Quant. Grav.",
	volume = "22",
	pages = "3315--3324",
	year = "2005"
}

@article{Mazza:2026taj,
	author = "Mazza, Jacopo",
	title = "{Black bounces to traversable wormholes from pure gravity in four and higher dimensions}",
	eprint = "2608.02771",
	archivePrefix = "arXiv",
	primaryClass = "gr-qc",
	month = "8",
	year = "2026"
}

@article{Colleaux:2026hat,
	author = "Coll{\'e}aux, Aimeric",
	title = "{Rational regular black holes in non-polynomial gravity}",
	eprint = "2608.17158",
	archivePrefix = "arXiv",
	primaryClass = "gr-qc",
	month = "8",
	year = "2026"
}

@article{Bueno:2025tli,
	author = "Bueno, Pablo and Hennigar, Robie A. and Murcia, {\'A}ngel J. and Vicente-Cano, Aitor",
	title = "{Buchdahl limits in theories with regular black holes}",
	eprint = "2512.19796",
	archivePrefix = "arXiv",
	primaryClass = "gr-qc",
	doi = "10.1103/mcrq-6fl3",
	journal = "Phys. Rev. D",
	volume = "113",
	number = "8",
	pages = "084008",
	year = "2026"
}

@article{Romans1992,
	author  = {Romans, L. J.},
	title   = {Supersymmetric, cold and lukewarm black holes in cosmological Einstein-Maxwell theory},
	journal = {Nuclear Physics B},
	volume  = {383},
	pages   = {395--415},
	year    = {1992},
	doi     = {10.1016/0550-3213(92)90684-4}
}

@article{MannRoss1995,
	author  = {Mann, R. B. and Ross, S. F.},
	title   = {Cosmological production of charged black hole pairs},
	journal = {Physical Review D},
	volume  = {52},
	pages   = {2254--2265},
	year    = {1995},
	doi     = {10.1103/PhysRevD.52.2254}
}

@article{Castro:2022cuo,
	author = "Castro, Alejandra and Mariani, Francesca and Toldo, Chiara",
	title = "{Near-extremal limits of de Sitter black holes}",
	eprint = "2212.14356",
	archivePrefix = "arXiv",
	primaryClass = "hep-th",
	doi = "10.1007/JHEP07(2023)131",
	journal = "JHEP",
	volume = "07",
	pages = "131",
	year = "2023"
}

@article{Yang:2021civ,
	author = "Yang, Run-Qiu and Cai, Rong-Gen and Li, Li",
	title = "{Constraining the number of horizons with energy conditions}",
	eprint = "2104.03012",
	archivePrefix = "arXiv",
	primaryClass = "gr-qc",
	doi = "10.1088/1361-6382/ac4118",
	journal = "Class. Quant. Grav.",
	volume = "39",
	number = "3",
	pages = "035005",
	year = "2022"
}

@article{Ling:2025ncw,
	author = "Ling, Yi and Yu, Zhangping",
	title = "{Big bounce and black bounce in quasi-topological gravity}",
	eprint = "2509.00137",
	archivePrefix = "arXiv",
	primaryClass = "gr-qc",
	doi = "10.1088/1475-7516/2026/03/004",
	journal = "JCAP",
	volume = "03",
	pages = "004",
	year = "2026"
}

@article{Einstein:1917ce,
	author = {Einstein, Albert},
	title = {Kosmologische Betrachtungen zur allgemeinen Relativit{\"a}tstheorie},
	journal = {Sitzungsberichte der K{\"o}niglich Preussischen Akademie der Wissenschaften},
	year = {1917},
	pages = {142--152}
}

@article{ORaifeartaigh:2017uct,
	author = "O'Raifeartaigh, Cormac and O'Keeffe, Michael and Nahm, Werner and Mitton, Simon",
	title = "{Einstein{\textquoteright}s 1917 static model of the universe: a centennial review}",
	eprint = "1701.07261",
	archivePrefix = "arXiv",
	primaryClass = "physics.hist-ph",
	doi = "10.1140/epjh/e2017-80002-5",
	journal = "Eur. Phys. J. H",
	volume = "42",
	number = "3",
	pages = "431--474",
	year = "2017"
}

@article{Hod:2024shg,
	author = "Hod, Shahar",
	title = "{Super-extremal black holes in the quasitopological electromagnetic field theory}",
	eprint = "2404.03744",
	archivePrefix = "arXiv",
	primaryClass = "gr-qc",
	doi = "10.1140/epjc/s10052-024-12454-w",
	journal = "Eur. Phys. J. C",
	volume = "84",
	number = "2",
	pages = "111",
	year = "2024"
}

@article{Camanho:2011rj,
	author = "Camanho, Xian O. and Edelstein, Jose D.",
	title = "{A Lovelock black hole bestiary}",
	eprint = "1103.3669",
	archivePrefix = "arXiv",
	primaryClass = "hep-th",
	doi = "10.1088/0264-9381/30/3/035009",
	journal = "Class. Quant. Grav.",
	volume = "30",
	pages = "035009",
	year = "2013"
}

@phdthesis{Camanho:2013pzg,
	author = "Camanho, Xi{\'a}n O.",
	title = "{Lovelock gravity, black holes and holography}",
	eprint = "1509.08129",
	archivePrefix = "arXiv",
	primaryClass = "hep-th",
	school = "Santiago de Compostela U.",
	year = "2013"
}

@article{Fernandes:2025fnz,
	author = "Fernandes, Pedro G. S.",
	title = "{Singularity resolution and inflation from an infinite tower of regularized curvature corrections}",
	eprint = "2504.07692",
	archivePrefix = "arXiv",
	primaryClass = "gr-qc",
	doi = "10.1103/763p-htct",
	journal = "Phys. Rev. D",
	volume = "112",
	number = "8",
	pages = "084028",
	year = "2025"
}

@article{Greenwood:2008ht,
	author = "Greenwood, Eric and Stojkovic, Dejan",
	title = "{Quantum gravitational collapse: Non-singularity and non-locality}",
	eprint = "0802.4087",
	archivePrefix = "arXiv",
	primaryClass = "gr-qc",
	doi = "10.1088/1126-6708/2008/06/042",
	journal = "JHEP",
	volume = "06",
	pages = "042",
	year = "2008"
}

@article{Bueno:2025dqk,
	author = "Bueno, Pablo and Lasso Andino, Oscar and Moreno, Javier and van der Velde, Guido",
	title = "{On regular charged black holes in three dimensions}",
	eprint = "2503.02930",
	archivePrefix = "arXiv",
	primaryClass = "gr-qc",
	doi = "10.1007/JHEP08(2025)132",
	journal = "JHEP",
	volume = "25",
	pages = "132",
	year = "2025"
}

@article{Harada:2025cwd,
	author = "Harada, Tomohiro and Chen, Chiang-Mei and Mandal, Rituparna",
	title = "{Singularity resolution and regular black hole formation in gravitational collapse in asymptotically safe gravity}",
	eprint = "2502.16787",
	archivePrefix = "arXiv",
	primaryClass = "gr-qc",
	reportNumber = "RUP-25-3",
	doi = "10.1103/pt9s-jqjz",
	journal = "Phys. Rev. D",
	volume = "111",
	number = "12",
	pages = "126017",
	year = "2025"
}

@article{Kawai:2017txu,
	author = "Kawai, Hikaru and Yokokura, Yuki",
	title = "{A Model of Black Hole Evaporation and 4D Weyl Anomaly}",
	eprint = "1701.03455",
	archivePrefix = "arXiv",
	primaryClass = "hep-th",
	doi = "10.3390/universe3020051",
	journal = "Universe",
	volume = "3",
	number = "2",
	pages = "51",
	year = "2017"
}

@article{Saini:2014qpa,
	author = "Saini, Anshul and Stojkovic, Dejan",
	title = "{Nonlocal (but also nonsingular) physics at the last stages of gravitational collapse}",
	eprint = "1401.6182",
	archivePrefix = "arXiv",
	primaryClass = "gr-qc",
	doi = "10.1103/PhysRevD.89.044003",
	journal = "Phys. Rev. D",
	volume = "89",
	number = "4",
	pages = "044003",
	year = "2014"
}

@article{Bonanno:2000ep,
	author = "Bonanno, Alfio and Reuter, Martin",
	title = "{Renormalization group improved black hole space-times}",
	eprint = "hep-th/0002196",
	archivePrefix = "arXiv",
	reportNumber = "INFN-CT-03-00, MZ-TH-00-04",
	doi = "10.1103/PhysRevD.62.043008",
	journal = "Phys. Rev. D",
	volume = "62",
	pages = "043008",
	year = "2000"
}

@article{Frolov:2014jva,
	author = "Frolov, Valeri P.",
	title = "{Information loss problem and a 'black hole` model with a closed apparent horizon}",
	eprint = "1402.5446",
	archivePrefix = "arXiv",
	primaryClass = "hep-th",
	doi = "10.1007/JHEP05(2014)049",
	journal = "JHEP",
	volume = "05",
	pages = "049",
	year = "2014"
}

@article{Zhang:2024ljd,
	author = "Zhang, Xiao-Yu and Zhao, Li and Wang, Yong-Qiang",
	title = "{Bardeen-Dirac stars in Anti-de Sitter spacetime}",
	eprint = "2409.14402",
	archivePrefix = "arXiv",
	primaryClass = "gr-qc",
	doi = "10.1088/1475-7516/2025/01/117",
	journal = "JCAP",
	volume = "01",
	pages = "117",
	year = "2025"
}

@article{Li:2024rbw,
	author = {Li, Zhi-Chao and L{\"u}, Hong},
	title = "{Regular electric black holes from Einstein-Maxwell-scalar gravity}",
	eprint = "2407.07952",
	archivePrefix = "arXiv",
	primaryClass = "gr-qc",
	doi = "10.1103/PhysRevD.110.104046",
	journal = "Phys. Rev. D",
	volume = "110",
	number = "10",
	pages = "104046",
	year = "2024"
}

@article{Boyanov:2025pes,
	author = "Boyanov, Valentin and Carballo-Rubio, Ra{\'u}l",
	title = "{Regular Vaidya Solutions of Effective Gravitational Theories}",
	eprint = "2506.14875",
	archivePrefix = "arXiv",
	primaryClass = "gr-qc",
	doi = "10.1103/3ng1-y6vs",
	journal = "Phys. Rev. Lett.",
	volume = "136",
	number = "17",
	pages = "171403",
	year = "2026"
}

@article{Konoplya:2025uta,
	author = "Konoplya, Roman. A. and Zhidenko, Alexander",
	title = "{Convergence of Higher-Curvature Expansions Near the Horizon: Hawking Radiation from Regular Black Holes}",
	eprint = "2507.22660",
	archivePrefix = "arXiv",
	primaryClass = "gr-qc",
	doi = "10.53941/ijgtp.2025.100005",
	journal = "Int. J. Grav. Theor. Phys.",
	volume = "1",
	number = "1",
	pages = "5",
	year = "2025"
}

@article{Cisterna:2025vxk,
	author = "Cisterna, Adolfo and Hassaine, Mokhtar and Hernandez-Vera, Ulises",
	title = "{Thermodynamics of four-dimensional regular black holes with an infinite tower of regularized curvature corrections}",
	eprint = "2505.23467",
	archivePrefix = "arXiv",
	primaryClass = "gr-qc",
	doi = "10.1103/6f3b-8794",
	journal = "Phys. Rev. D",
	volume = "112",
	number = "6",
	pages = "064036",
	year = "2025"
}

@article{Skvortsova:2024wly,
	author = "Skvortsova, Milena",
	title = "{Ringing of Extreme Regular Black Holes}",
	eprint = "2405.15807",
	archivePrefix = "arXiv",
	primaryClass = "gr-qc",
	doi = "10.1134/S020228932470018X",
	journal = "Grav. Cosmol.",
	volume = "30",
	number = "3",
	pages = "279--288",
	year = "2024"
}

@article{Bolokhov:2024sdy,
	author = "Bolokhov, S. V. and Bronnikov, K. A. and Skvortsova, M. V.",
	title = "{A Regular Center Instead of a Black Bounce}",
	eprint = "2405.09124",
	archivePrefix = "arXiv",
	primaryClass = "gr-qc",
	doi = "10.1134/S0202289324700178",
	journal = "Grav. Cosmol.",
	volume = "30",
	number = "3",
	pages = "265--278",
	year = "2024"
}

@article{Bronnikov:2024izh,
	author = "Bronnikov, Kirill A.",
	title = "{Regular black holes as an alternative to black bounce}",
	eprint = "2404.14816",
	archivePrefix = "arXiv",
	primaryClass = "gr-qc",
	doi = "10.1103/PhysRevD.110.024021",
	journal = "Phys. Rev. D",
	volume = "110",
	number = "2",
	pages = "024021",
	year = "2024"
}

@article{Alencar:2024yvh,
	author = "Alencar, G. and Bronnikov, Kirill A. and Rodrigues, Manuel E. and S\'aez-Chill\'on G\'omez, Diego and de S. Silva, Marcos V.",
	title = "{On black bounce space-times in non-linear electrodynamics}",
	eprint = "2403.12897",
	archivePrefix = "arXiv",
	primaryClass = "gr-qc",
	doi = "10.1140/epjc/s10052-024-13119-4",
	journal = "Eur. Phys. J. C",
	volume = "84",
	number = "7",
	pages = "745",
	year = "2024"
}

@article{Ditta:2024iky,
	author = "Ditta, Allah and Bouzenada, Abdelmalek and Mustafa, G. and Alanazi, Yousef Mohammed and Mushtaq, Farzan",
	title = "{Particle motion, shadows and thermodynamics of regular black hole in pure gravity}",
	doi = "10.1016/j.dark.2024.101573",
	journal = "Phys. Dark Univ.",
	volume = "46",
	pages = "101573",
	year = "2024"
}

@article{Cisterna:2024ksz,
	author = "Cisterna, Adolfo and Grandi, Nicolas and Oliva, Julio",
	title = "{de Sitter geometric inflation from dynamical singularities}",
	eprint = "2406.10037",
	archivePrefix = "arXiv",
	primaryClass = "hep-th",
	doi = "10.1103/PhysRevD.110.084043",
	journal = "Phys. Rev. D",
	volume = "110",
	number = "8",
	pages = "084043",
	year = "2024"
}

@article{Junior:2023ixh,
	author = "Junior, Jos\'e Tarciso S. S. and Lobo, Francisco S. N. and Rodrigues, Manuel E.",
	title = "{(Regular) Black holes in conformal Killing gravity coupled to nonlinear electrodynamics and scalar fields}",
	eprint = "2310.19508",
	archivePrefix = "arXiv",
	primaryClass = "gr-qc",
	doi = "10.1088/1361-6382/ad210e",
	journal = "Class. Quant. Grav.",
	volume = "41",
	number = "5",
	pages = "055012",
	year = "2024"
}

@article{Frolov:1989pf,
	author = "Frolov, Valeri P. and Markov, M. A. and Mukhanov, Viatcheslav F.",
	editor = "Markov, Moisei Aleksandrovich and Berezin, Viktor A. and Mukhanov, Viatcheslav F. F.",
	title = "{THROUGH A BLACK HOLE INTO A NEW UNIVERSE?}",
	doi = "10.1016/0370-2693(89)91114-3",
	journal = "Phys. Lett. B",
	volume = "216",
	pages = "272--276",
	year = "1989"
}

@article{Barrabes:1995nk,
	author = "Barrabes, Claude and Frolov, Valeri P.",
	title = "{How many new worlds are inside a black hole?}",
	eprint = "hep-th/9511136",
	archivePrefix = "arXiv",
	reportNumber = "ALBERTA-THY-30-95",
	doi = "10.1103/PhysRevD.53.3215",
	journal = "Phys. Rev. D",
	volume = "53",
	pages = "3215--3223",
	year = "1996"
}

@article{Nicolini:2005vd,
	author = "Nicolini, Piero and Smailagic, Anais and Spallucci, Euro",
	title = "{Noncommutative geometry inspired Schwarzschild black hole}",
	eprint = "gr-qc/0510112",
	archivePrefix = "arXiv",
	doi = "10.1016/j.physletb.2005.11.004",
	journal = "Phys. Lett. B",
	volume = "632",
	pages = "547--551",
	year = "2006"
}

@article{Olmo:2012nx,
	author = "Olmo, Gonzalo J. and Rubiera-Garcia, D.",
	title = {{Reissner-Nordstr\"om black holes in extended Palatini theories}},
	eprint = "1207.6004",
	archivePrefix = "arXiv",
	primaryClass = "gr-qc",
	doi = "10.1103/PhysRevD.86.044014",
	journal = "Phys. Rev. D",
	volume = "86",
	pages = "044014",
	year = "2012"
}

@article{Balakin:2015gpq,
	author = "Balakin, Alexander B. and Lemos, Jos\'e P. S. and Zayats, Alexei E.",
	title = "{Regular nonminimal magnetic black holes in spacetimes with a cosmological constant}",
	eprint = "1512.02653",
	archivePrefix = "arXiv",
	primaryClass = "gr-qc",
	doi = "10.1103/PhysRevD.93.024008",
	journal = "Phys. Rev. D",
	volume = "93",
	number = "2",
	pages = "024008",
	year = "2016"
}

@article{Bazeia:2015uia,
	author = "Bazeia, D. and Losano, L. and Olmo, Gonzalo J. and Rubiera-Garcia, D. and Sanchez-Puente, A.",
	title = "{Classical resolution of black hole singularities in arbitrary dimension}",
	eprint = "1507.07763",
	archivePrefix = "arXiv",
	primaryClass = "hep-th",
	reportNumber = "IFIC-15-52",
	doi = "10.1103/PhysRevD.92.044018",
	journal = "Phys. Rev. D",
	volume = "92",
	number = "4",
	pages = "044018",
	year = "2015"
}

@article{Chamseddine:2016ktu,
	author = "Chamseddine, Ali H. and Mukhanov, Viatcheslav",
	title = "{Nonsingular Black Hole}",
	eprint = "1612.05861",
	archivePrefix = "arXiv",
	primaryClass = "gr-qc",
	doi = "10.1140/epjc/s10052-017-4759-z",
	journal = "Eur. Phys. J. C",
	volume = "77",
	number = "3",
	pages = "183",
	year = "2017"
}

@article{Bambi:2016xme,
	author = "Bambi, Cosimo and Rubiera-Garcia, D. and Wang, Yixu",
	title = "{Black hole solutions in functional extensions of Born-Infeld gravity}",
	eprint = "1608.04873",
	archivePrefix = "arXiv",
	primaryClass = "gr-qc",
	doi = "10.1103/PhysRevD.94.064002",
	journal = "Phys. Rev. D",
	volume = "94",
	number = "6",
	pages = "064002",
	year = "2016"
}

@article{Bejarano:2017fgz,
	author = "Bejarano, Cecilia and Olmo, Gonzalo J. and Rubiera-Garcia, Diego",
	title = "{What is a singular black hole beyond General Relativity?}",
	eprint = "1702.01292",
	archivePrefix = "arXiv",
	primaryClass = "hep-th",
	reportNumber = "IFIC-17-04",
	doi = "10.1103/PhysRevD.95.064043",
	journal = "Phys. Rev. D",
	volume = "95",
	number = "6",
	pages = "064043",
	year = "2017"
}

@article{Colleaux:2017ibe,
	author = "Coll\'eaux, Aimeric and Chinaglia, Stefano and Zerbini, Sergio",
	title = "{Nonpolynomial Lagrangian approach to regular black holes}",
	eprint = "1712.03730",
	archivePrefix = "arXiv",
	primaryClass = "gr-qc",
	doi = "10.1142/S0218271818300021",
	journal = "Int. J. Mod. Phys. D",
	volume = "27",
	number = "03",
	pages = "1830002",
	year = "2018"
}

@article{Cano:2018aod,
	author = "Cano, Pablo A. and Chimento, Samuele and Ort\'\i{}n, Tom\'as and Ruip\'erez, Alejandro",
	title = "{Regular Stringy Black Holes?}",
	eprint = "1806.08377",
	archivePrefix = "arXiv",
	primaryClass = "hep-th",
	reportNumber = "IFT-UAM-CSIC-18-62, IFT-UAM/CSIC-18-62",
	doi = "10.1103/PhysRevD.99.046014",
	journal = "Phys. Rev. D",
	volume = "99",
	number = "4",
	pages = "046014",
	year = "2019"
}

@phdthesis{Colleaux:2019ckh,
	author = "Colleaux, Aimeric",
	title = "{Regular black hole and cosmological spacetimes in Non-Polynomial Gravity theories}",
	school = "Trento U.",
	month = "6",
	year = "2019"
}

@article{Guerrero:2020uhn,
	author = "Guerrero, Merce and Rubiera-Garcia, Diego",
	title = "{Nonsingular black holes in nonlinear gravity coupled to Euler-Heisenberg electrodynamics}",
	eprint = "2005.08828",
	archivePrefix = "arXiv",
	primaryClass = "gr-qc",
	doi = "10.1103/PhysRevD.102.024005",
	journal = "Phys. Rev. D",
	volume = "102",
	number = "2",
	pages = "024005",
	year = "2020"
}

@article{Brandenberger:2021jqs,
	author = "Brandenberger, Robert and Heisenberg, Lavinia and Robnik, Jakob",
	title = "{Through a black hole into a new universe}",
	eprint = "2105.07166",
	archivePrefix = "arXiv",
	primaryClass = "hep-th",
	doi = "10.1142/S0218271821420013",
	journal = "Int. J. Mod. Phys. D",
	volume = "30",
	number = "14",
	pages = "2142001",
	year = "2021"
}

@article{Olmo:2022cui,
	author = "Olmo, Gonzalo J. and Rubiera-Garcia, Diego",
	title = "{Regular black holes in Palatini gravity}",
	eprint = "2209.05061",
	archivePrefix = "arXiv",
	primaryClass = "gr-qc",
	month = "9",
	year = "2022"
}

@article{Junior:2024xmm,
	author = "Junior, Jos{\'e} Tarciso S. S. and Lobo, Francisco S. N. and Rodrigues, Manuel E.",
	title = "{Black holes and regular black holes in coincident $f({\mathbb {Q}},{\mathbb {B}}_Q)$ gravity coupled to nonlinear electrodynamics}",
	eprint = "2402.02534",
	archivePrefix = "arXiv",
	primaryClass = "gr-qc",
	doi = "10.1140/epjc/s10052-024-12696-8",
	journal = "Eur. Phys. J. C",
	volume = "84",
	number = "3",
	pages = "332",
	year = "2024"
}

@article{Tan:2025hht,
	author = "Tan, Chen and Wang, Yong-Qiang",
	title = "{Frozen Neutron Stars in Four-Dimensional Non-polynomial Gravities}",
	eprint = "2512.23525",
	archivePrefix = "arXiv",
	primaryClass = "gr-qc",
	month = "12",
	year = "2025"
}

@article{Chen:2025iuy,
	author = "Chen, Jun-Ru and Wang, Yong-Qiang",
	title = "{Proca stars and their frozen states in an infinite tower of higher-derivative gravity}",
	eprint = "2512.24584",
	archivePrefix = "arXiv",
	primaryClass = "gr-qc",
	month = "12",
	year = "2025"
}

@article{Konoplya:2026gim,
	author = "Konoplya, R. A.",
	title = "{Quasinormal modes of four-dimensional regular black holes in quasi-topological gravity: Overtones{\textquoteright} outburst via WKB method}",
	eprint = "2603.03189",
	archivePrefix = "arXiv",
	primaryClass = "gr-qc",
	doi = "10.1016/j.physletb.2026.140386",
	journal = "Phys. Lett. B",
	volume = "876",
	pages = "140386",
	year = "2026"
}

@article{Li:2026mam,
	author = "Li, Gen and Wang, Yong-Qiang",
	title = "{Ellis--Bronnikov wormhole in Quasi-topological Gravity}",
	eprint = "2602.01029",
	archivePrefix = "arXiv",
	primaryClass = "gr-qc",
	month = "2",
	year = "2026"
}

@article{Arbelaez:2026eaz,
	author = "Arbelaez, Juan Pablo",
	title = "{Grey-body factors of higher dimensional regular black holes in quasi-topological theories}",
	eprint = "2601.22340",
	archivePrefix = "arXiv",
	primaryClass = "gr-qc",
	doi = "10.1088/1475-7516/2026/05/043",
	journal = "JCAP",
	volume = "05",
	pages = "043",
	year = "2026"
}

@article{Aguayo:2025xfi,
	author = "Aguayo, Monserrat and Gajardo, Leonardo and Grandi, Nicol{\'a}s and Moreno, Javier and Oliva, Julio and Reyes, Mart{\'\i}n",
	title = "{Holographic explorations of regular black holes in pure gravity}",
	eprint = "2505.11736",
	archivePrefix = "arXiv",
	primaryClass = "hep-th",
	doi = "10.1007/JHEP09(2025)030",
	journal = "JHEP",
	volume = "09",
	pages = "030",
	year = "2025"
}

@article{Fernandes:2025eoc,
	author = "Fernandes, Pedro G. S.",
	title = "{Regular BTZ black holes from an infinite tower of corrections}",
	eprint = "2504.08565",
	archivePrefix = "arXiv",
	primaryClass = "gr-qc",
	doi = "10.1016/j.physletb.2025.139772",
	journal = "Phys. Lett. B",
	volume = "868",
	pages = "139772",
	year = "2025"
}

@article{Eichhorn:2025pgy,
	author = "Eichhorn, Astrid and Fernandes, Pedro G. S.",
	title = "{Regular black holes without mass-inflation instability and gravastars from modified gravity}",
	eprint = "2508.00686",
	archivePrefix = "arXiv",
	primaryClass = "gr-qc",
	doi = "10.1103/nqz2-88zf",
	journal = "Phys. Rev. D",
	volume = "113",
	number = "8",
	pages = "L081501",
	year = "2026"
}

@article{Fernandes:2025mic,
	author = "Fernandes, Pedro G. S. and Gou, Jingqian and Heisenberg, Lavinia and Nussbaumer, Nadine",
	title = "{Inflation, black holes with primary hair, and regular planar black holes from an infinite tower of regularized Lovelock-Proca corrections}",
	eprint = "2511.22798",
	archivePrefix = "arXiv",
	primaryClass = "gr-qc",
	doi = "10.1088/1475-7516/2026/04/078",
	journal = "JCAP",
	volume = "04",
	pages = "078",
	year = "2026"
}

@article{Hao:2025utc,
	author = "Hao, Chen-Hao and Jing, Jiliang and Wang, Jieci",
	title = "{Charged regular black holes from quasi-topological gravities in D {\ensuremath{\geq}} 5}",
	eprint = "2512.04604",
	archivePrefix = "arXiv",
	primaryClass = "gr-qc",
	doi = "10.1088/1475-7516/2026/05/067",
	journal = "JCAP",
	volume = "05",
	pages = "067",
	year = "2026"
}

@article{Frolov:2025ddw,
	author = "Frolov, Valeri P.",
	title = "{Quasitopological gravity and double-copy formalism}",
	eprint = "2512.14674",
	archivePrefix = "arXiv",
	primaryClass = "gr-qc",
	doi = "10.1103/mbdj-tqtv",
	journal = "Phys. Rev. D",
	volume = "113",
	number = "6",
	pages = "064023",
	year = "2026"
}

@article{Frolov:2026tft,
	author = "Frolov, Valeri P. and Yoo, Chul-Moon and Zelnikov, Andrei",
	title = "{Vaidya-Type Solutions of Quasitopological Gravity Interacting with Nonlinear Electrodynamics}",
	eprint = "2607.22856",
	archivePrefix = "arXiv",
	primaryClass = "gr-qc",
	reportNumber = "Alberta Thy 1-26, NU-QG-23",
	month = "7",
	year = "2026"
}

@article{Fan:2026mdu,
	author = "Fan, Sihao and Wu, Chen and Guo, Wenjun",
	title = "{Black hole correspondences of quasi-topological gravity : Shadows, quasinormal modes, graybody factors}",
	eprint = "2608.30169",
	archivePrefix = "arXiv",
	primaryClass = "gr-qc",
	month = "8",
	year = "2026"
}

@article{Frolov:2026byy,
	author = "Frolov, Valeri P.",
	title = "{Quasitopological Gravity with Matter: Modified Double-Copy Approach}",
	eprint = "2608.12596",
	archivePrefix = "arXiv",
	primaryClass = "gr-qc",
	month = "8",
	year = "2026"
}

@article{Bueno:2026bby,
	author = "Bueno, Pablo and Cano, Pablo A. and Hennigar, Robie A. and Moreno, Javier and Murcia, {\'A}ngel J. and van der Velde, Guido",
	title = "{Logarithmic correction to the entropy of near-extremal higher-curvature black holes}",
	eprint = "2609.07801",
	archivePrefix = "arXiv",
	primaryClass = "hep-th",
	reportNumber = "YITP-26-120",
	month = "9",
	year = "2026"
}

@article{Borissova:2025hmj,
	author = "Borissova, Johanna and Liberati, Stefano and Visser, Matt",
	title = "{Timelike convergence condition in regular black-hole spacetimes with (anti{\textendash})de Sitter core}",
	eprint = "2509.08590",
	archivePrefix = "arXiv",
	primaryClass = "gr-qc",
	doi = "10.1103/rrc9-g1sv",
	journal = "Phys. Rev. D",
	volume = "112",
	number = "10",
	pages = "104072",
	year = "2025"
}

@article{Borissova:2026prd,
	author = "Borissova, Johanna",
	title = "{Formation of extremal regular black holes}",
	eprint = "2608.02908",
	archivePrefix = "arXiv",
	primaryClass = "gr-qc",
	reportNumber = "Imperial/TP/2026/JB/07",
	month = "8",
	year = "2026"
}

@article{Borissova:2026rbi,
	author = "Borissova, Johanna",
	title = "{$g_{tt}g_{rr} =-1$ black hole thermodynamics in extended quasi-topological gravity}",
	eprint = "2604.24101",
	archivePrefix = "arXiv",
	primaryClass = "gr-qc",
	reportNumber = "Imperial/TP/2026/JB/05",
	doi = "10.1103/gbdk-xm4c",
	journal = "Phys. Rev. D",
	volume = "114",
	number = "6",
	pages = "064038",
	year = "2026"
}

@article{Zaslavskii:2010qz,
	author = "Zaslavskii, O. B.",
	title = "{Regular black holes and energy conditions}",
	eprint = "1004.2362",
	archivePrefix = "arXiv",
	primaryClass = "gr-qc",
	doi = "10.1016/j.physletb.2010.04.031",
	journal = "Phys. Lett. B",
	volume = "688",
	pages = "278--280",
	year = "2010"
}

@article{Maeda:2018hqu,
	author = "Maeda, Hideki and Martinez, Cristian",
	title = "{Energy conditions in arbitrary dimensions}",
	eprint = "1810.02487",
	archivePrefix = "arXiv",
	primaryClass = "gr-qc",
	doi = "10.1093/ptep/ptaa009",
	journal = "PTEP",
	volume = "2020",
	number = "4",
	pages = "043E02",
	year = "2020"
}

@article{Lan:2022bld,
	author = "Lan, Chen and Miao, Yan-Gang and Zang, Yi-Xiong",
	title = "{Regular black holes with improved energy conditions and their analogues in fluids*}",
	eprint = "2206.08694",
	archivePrefix = "arXiv",
	primaryClass = "gr-qc",
	doi = "10.1088/1674-1137/acc1cd",
	journal = "Chin. Phys. C",
	volume = "47",
	number = "5",
	pages = "052001",
	year = "2023"
}

@article{Ye:2026lah,
	author = "Ye, Jing-Peng and Yang, Run-Qiu",
	title = "{Constrain the number of inner horizons from gravitational Noether charges}",
	eprint = "2610.30812",
	archivePrefix = "arXiv",
	primaryClass = "gr-qc",
	month = "10",
	year = "2026"
}

@article{Borissova:2025msp,
	author = "Borissova, Johanna and Liberati, Stefano and Visser, Matt",
	title = "{Violations of the null convergence condition in kinematical transitions between singular and regular black holes, horizonless compact objects, and bounces}",
	eprint = "2502.00548",
	archivePrefix = "arXiv",
	primaryClass = "gr-qc",
	doi = "10.1103/PhysRevD.111.104054",
	journal = "Phys. Rev. D",
	volume = "111",
	number = "10",
	pages = "104054",
	year = "2025"
}

@article{Borissova:2026ohj,
	author = "Borissova, Johanna and Liberati, Stefano and Visser, Matt",
	title = "{Null geodesic defocusing in dynamical black-hole-to-white-hole transitions}",
	eprint = "2606.22404",
	archivePrefix = "arXiv",
	primaryClass = "gr-qc",
	reportNumber = "Imperial/TP/2026/JB/06",
	month = "6",
	year = "2026"
}
		
	\end{document}